\documentclass[11pt]{paper}
\usepackage[T1]{fontenc}
\usepackage[latin1]{inputenc}
\usepackage{geometry}
\usepackage{subfigure}
\usepackage{amsmath}
\usepackage{graphicx}
\usepackage{amssymb}
\usepackage[authoryear]{natbib}

\begin{document}

\title{Bistability of the \emph{lac} operon during growth of \emph{Escherichia
coli} on lactose and lactose + glucose}

\maketitle
\begin{center}Atul Narang%
\footnote{Email: narang@che.ufl.edu%
}\par\end{center}

\begin{center}Department of Chemical Engineering, University of Florida,
Gainesville, FL~32611-6005.\par\end{center}

\begin{center}Sergei S. Pilyugin\par\end{center}

\begin{center}Department of Mathematics, University of Florida, Gainesville,
FL~32611-8105.\par\end{center}

\begin{abstract}
The \emph{lac} operon of \emph{Escherichia coli} can exhibit bistability.
Early studies showed that bistability occurs during growth on TMG/succinate
and lactose + glucose, but not during growth on lactose. More recently,
studies with lacGFP-transfected cells show bistability during growth
on TMG/succinate, but not during growth on lactose and lactose + glucose.
In the literature, these results are invariably attributed to variations
in the destabilizing effect of the positive feedback generated by
induction. Specifically, during growth on TMG/succinate, \emph{lac}
induction generates strong positive feedback because the permease
stimulates the accumulation of intracellular TMG, which, in turn,
promotes the synthesis of even more permease. This positive feedback
is attenuated during growth on lactose because hydrolysis of intracellular
lactose by $\beta$-galactosidase suppresses the stimulatory effect
of the permease. It is attenuated even more during growth on lactose
+ glucose because glucose inhibits the uptake of lactose. But it is
clear that the stabilizing effect of dilution also changes dramatically
as a function of the medium composition. For instance, during growth
on TMG/succinate, the dilution rate of \emph{lac} permease is proportional
to its activity, $e$, because the specific growth rate is independent
of $e$ (it is completely determined by the concentration of succinate).
However, during growth on lactose, the dilution rate of the permease
is proportional to $e^{2}$ because the specific growth rate is proportional
to the specific lactose uptake rate, which in turn, proportional to
$e$. We show that: (a)~This dependence on $e^{2}$ creates such
a strong stabilizing effect that bistability is virtually impossible
during growth on lactose, even in the face of the intense positive
feedback generated by induction. (b)~This stabilizing effect is weakened
during growth on lactose + glucose because the specific growth rate
on glucose is independent of $e$, so that the dilution rate once
again contains a term that is proportional to $e$. These results
imply that the \emph{lac} operon is much more prone to bistability
if the medium contain carbon sources that cannot be metabolized by
the \emph{lac} enzymes, e.g., succinate during growth on TMG/succinate
and glucose during growth on lactose + glucose. We discuss the experimental
data in the light of these results.
\end{abstract}

\section{Introduction}

The \emph{lac} operon has been a topic of considerable interest since
the late 1940's. This interest was stimulated by the hope that insights
into the mechanism of \emph{lac} induction would shed light on the
central problem of development, namely, the mechanism by which genetically
identical cells acquire distinct phenotypes~\citep{monod47,Spiegelman1948}.

Many of the early studies were concerned with the kinetics of enzyme
induction. Initial attempts to measure the kinetics were hindered
by the fact that lactose, the substrate that stimulates the induction
of the \emph{lac} operon, promotes not only the synthesis of the \emph{lac}
enzymes, but also their dilution by growth. Under these conditions,
it is impossible to separate the kinetics of enzyme synthesis from
the masking effects of dilution. This obstacle was overcome by the
discovery of \emph{gratuitous inducers}, such as methyl galactoside
(MG) and thiomethyl galactoside (TMG). Enzyme synthesis and dilution
could be uncoupled by exposing the cells to a medium containing a
gratuituous inducer and non-galactosidic carbon sources, such as glucose
or/and succinate. The gratuitous inducer promoted enzyme synthesis,
but not growth, and the non-galactosidic carbon sources supported
growth, but not enzyme synthesis.

\begin{figure}
\subfigure[]{\includegraphics[width=3in,keepaspectratio]{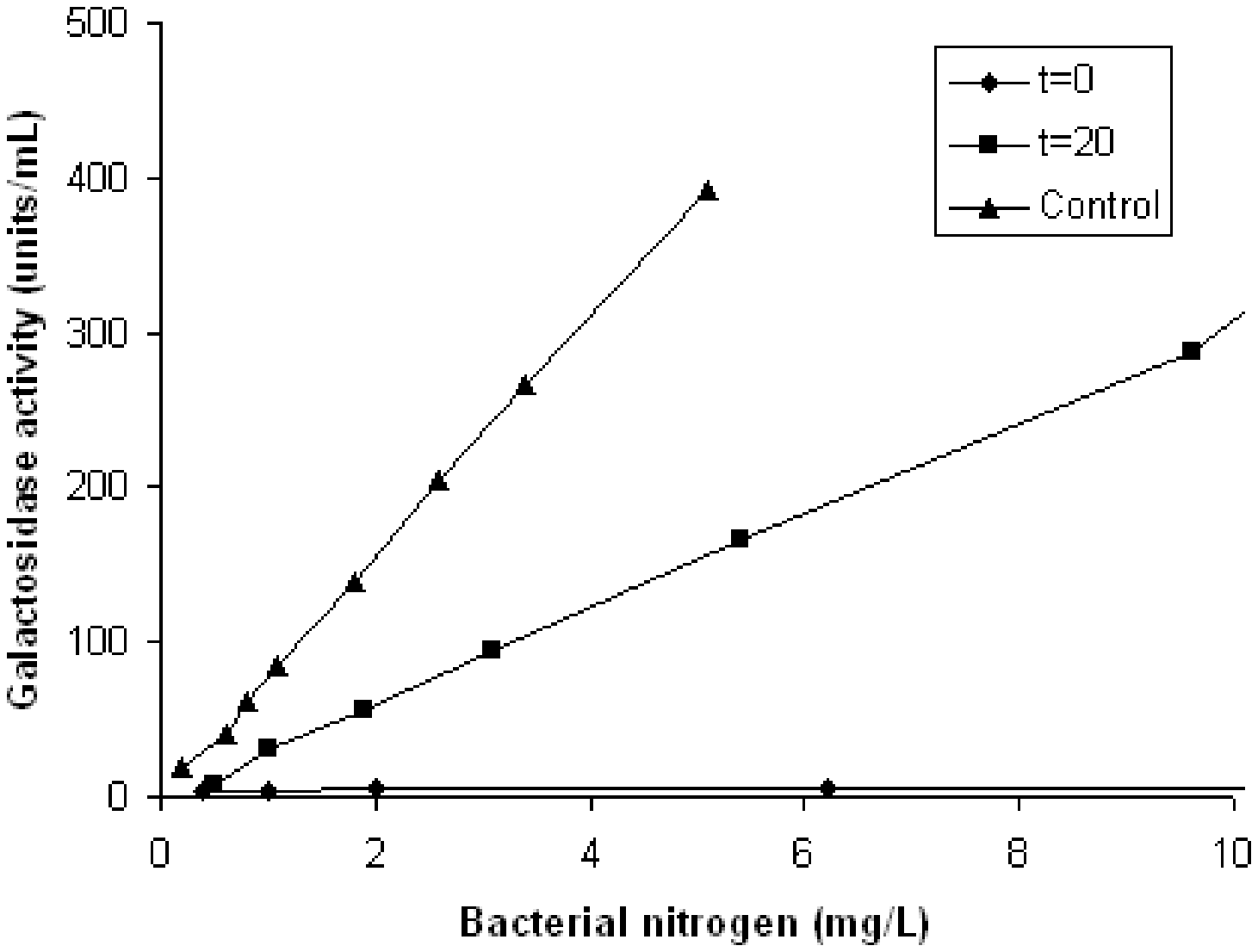}}\subfigure[]{\includegraphics[width=3in,keepaspectratio]{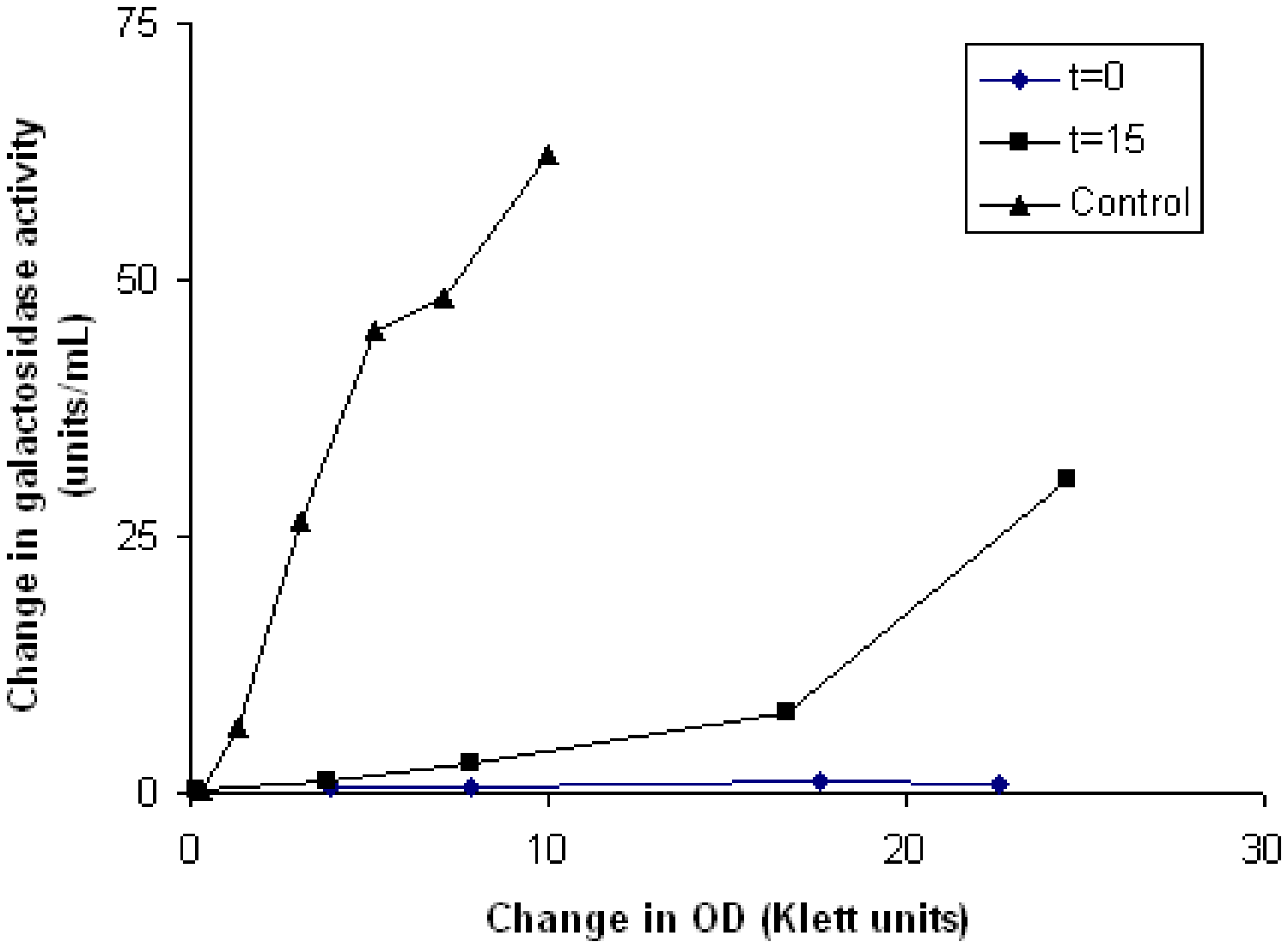}}

\caption{\label{f:CohnLoomis}Bistability during growth of \emph{E. coli}
on TMG~\citep[Fig.~4]{cohn59a} and lactose + glucose~\citep[Fig.~3]{Loomis1967}.
(a)~If glucose and TMG are added simultaneously to a culture growing
on succinate, there is no $\beta$-galactosidase synthesis ($\blacklozenge$).
If glucose is added to the culture 20 mins after the addition of TMG,
the enzyme is synthesized ($\blacksquare$) at a rate that is 50\%
of the rate observed when only TMG is added to the culture ($\blacktriangle$).
(b)~If lactose is added to a culture growing on glucose, there is
no $\beta$-galactosidase synthesis ($\blacklozenge$). If lactose
and glucose are added to the culture after it has been exposed to
IPTG for 15~mins, the enzyme synthesis rate ($\blacksquare$) increases
within a few generation to $\sim$40\% of the rate in a culture exposed
to IPTG only ($\blacktriangle$).}
\end{figure}

Although gratuitous inducers enabled enzyme synthesis and dilution
to be uncoupled, an important question remained. Specifically, it
was not known whether gratuitous inducers provoked the same enzyme
synthesis rate in every cell of a culture. Initial experiments suggested
that this was indeed the case. Benzer showed that~\citep[Figs.~6 and 7]{benzer53}:

\begin{enumerate}
\item If non-induced cells of \emph{E. coli} B (pregrown on lactate) were
exposed to 1~g/L of lactose, only a small fraction of the cells synthesized
$\beta$-galactosidase initially. This fraction increased progressively
until the culture became homogeneous eventually.
\item In sharp contrast, if non-induced cells were exposed to 2~g/L of
MG, all the cells started synthesizing $\beta$-galactosidase immediately
and at near-maximal rates, i.e., the population became homogeneous
almost instantly.
\end{enumerate}
However, it was shown later that the population became homogeneous
instantly only because the concentration of the gratuitous inducer
was high. At low concentrations of the gratuitious inducer:

\begin{enumerate}
\item The population remained heterogeneous for a significant period of
time. Furthermore, the smaller the concentration of the gratuitous
inducer, the longer the time required for the population to become
homogeneous~\citep[Table~1]{cohn59b}.
\item The enzyme synthesis rate was not uniquely determined by the composition
of the medium (Fig.~\ref{f:CohnLoomis}a). If TMG and glucose were
added simultaneously to a culture of \emph{E. coli} ML30 growing on
succinate, there was almost no synthesis of $\beta$-galactosidase.
However, if TMG was added 15~mins before the addition of glucose,
$\beta$-galactosidase was synthesized for up to 130 generations.
Thus, enzyme synthesis is bistable: Pre-induced cells remain induced,
and non-induced cells remain non-induced.
\end{enumerate}
The existence of bistability and heterogeneity depended crucially
upon the existence of \emph{lac} permease (LacY). Both phenomena disappeared
in \emph{lacY}$^{-}$ (cryptic), but not \emph{lac}Z$^{-}$, mutants~\citep{cohn59a}.

Subsequent experiments showed that bistability also occurred during
growth of \emph{E. coli} K12 3.000 on lactose + glucose~(Fig.~\ref{f:CohnLoomis}b).
If lactose was added to a culture growing on glucose, there was no
$\beta$-galactosidase synthesis. However, if the culture was exposed
to $10^{-3}$~M IPTG before the addition of lactose, $\beta$-galactosidase
synthesis persisted for several generations.

These intricate dynamics attracted significant attention among some
theoreticians~\citep[reviewed in][]{Laurent2005}. In particular,
Babloyantz \& Sanglier formulated a model of growth on TMG/succinate
which took due account of enzyme synthesis by the Jacob-Monod mechanism,
and enzyme depletion by degradation~\citep{babloyantz72}. They showed
that the model yielded the bistability observed in experiments. Chung
and Stephanopoulos formulated a similar model, the main differences
being that repressor-operator and repressor-inducer binding were assumed
to be in quasi-equilibrium, and the enzyme was depleted by both degradation
and dilution~\citep{chung96}. This model is given by the equations\begin{align}
\frac{dx}{dt} & =r_{s}-r_{x}^{-}-r_{g}x,\; r_{s}\equiv V_{s}e\frac{s}{K_{s}+s},\; r_{x}^{-}\equiv k_{x}^{-}x.\label{eq:ChungX}\\
\frac{de}{dt} & =r_{e}^{+}-r_{e}^{-}-r_{g}e,\; r_{e}^{+}\equiv V_{e}\frac{1+K_{x}^{2}x^{2}}{1+\alpha+K_{x}^{2}x^{2}},\; r_{e}^{-}\equiv k_{e}^{-}e\label{eq:ChungE}\end{align}
where $x$ and $s$ denote the intracellular and extracellular TMG
concentrations, respectively; $e$ denotes the \emph{lac} permease
activity; $r_{g}$ is the specific growth rate on the non-galactosidic
carbon source; $r_{s},r_{x}^{-}$ denote the specific rates of TMG
uptake and expulsion, respectively; and $r_{e}^{+},r_{e}^{-}$ denote
the specific rates of permease synthesis and degradation, respectively.
The expression for $r_{e}^{+}$ is based on the molecular model formulated
by Yagil \& Yagil, which assumes that the \emph{lac} operon contains
one operator, and the \emph{lac} repressor contains identical two
inducer-binding sites~\citep{yagil71}. The parameter, $K_{x}$,
is the association constant for the repressor-inducer binding; and
$\alpha$ is jointly proportional to the intracellular repressor level
and the association constant for repressor-operator binding. Evidently,
$\alpha$ is a measure of the \emph{repression}, defined as the ratio,
$\left.r_{e}^{+}\right|_{x\rightarrow\infty}/\left.r_{e}^{+}\right|_{x=0}$.

\begin{figure}[t]
\noindent \begin{centering}\includegraphics[width=2.5in,keepaspectratio]{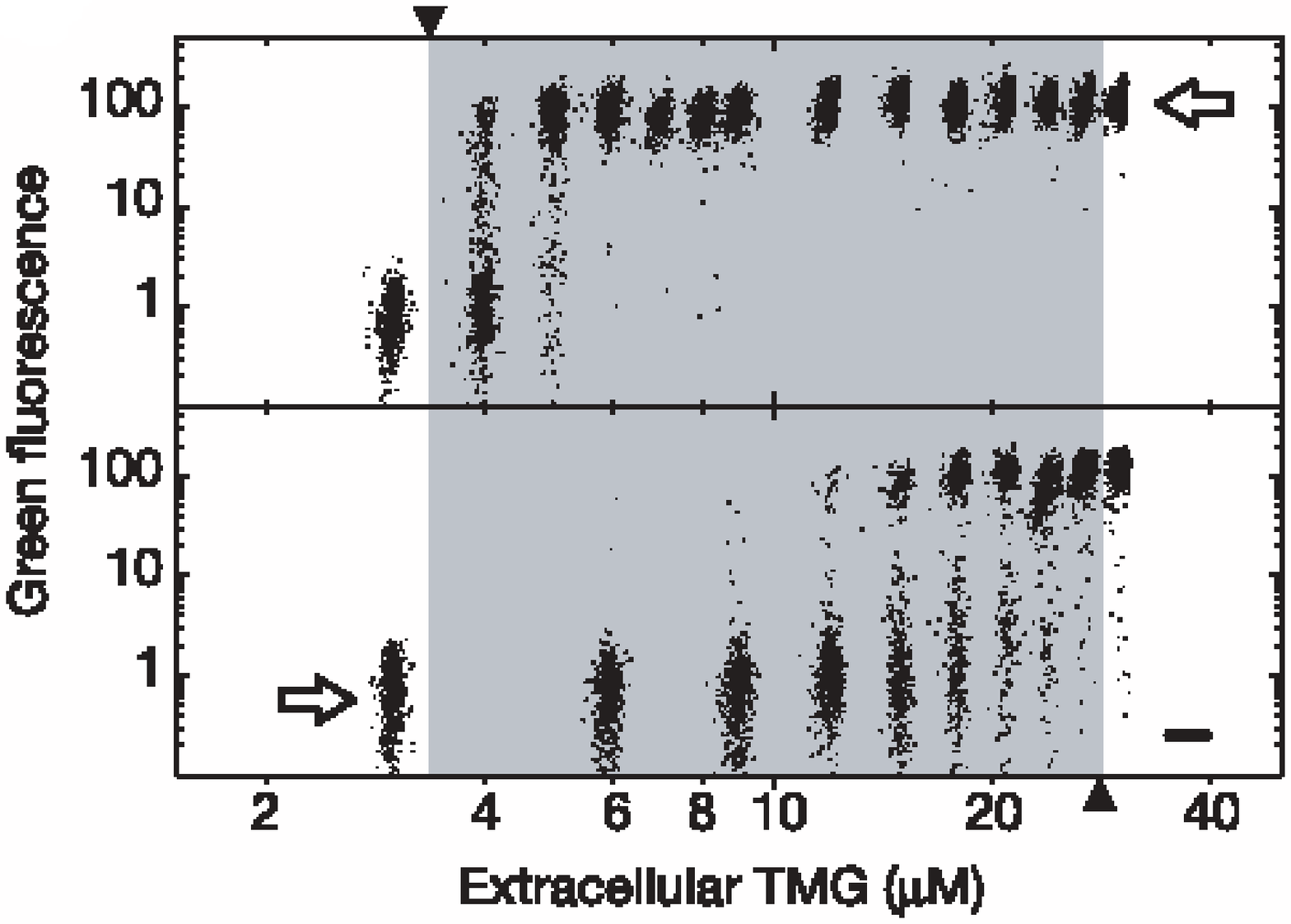}\hspace*{0.2in}\includegraphics[width=2.5in,keepaspectratio]{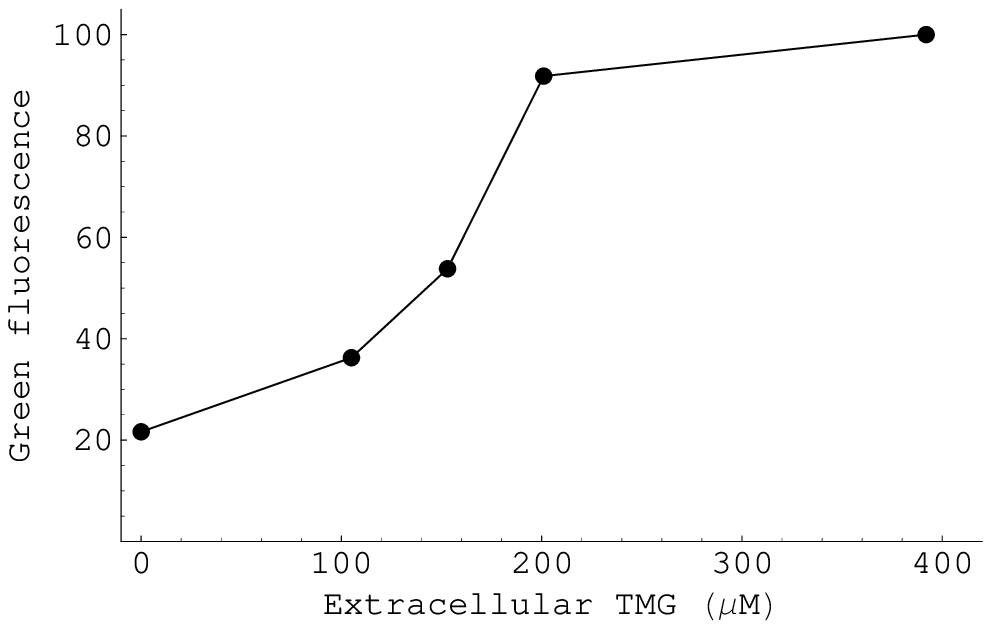}\par\end{centering}

\caption{\label{f:Oudenaarden}Dynamics of the \emph{lac} operon~\citep[from][]{ozbudak04}.
(a)~Bistability in wild-type cells. The (normalized) green fluorescence
provides a measure of the steady state activity of the \emph{lac}
operon. The upper (resp., lower) panel shows the green fluorescence
observed when an induced (resp., non-induced) inoculum of \emph{Escherichia
coli} is grown exponentially on a mixture of succinate and various
concentrations of extracellular TMG. Bistability occurs at TMG concentrations
between 3 and 30~$\mu$M: The green fluorescence is $\sim$100 if
the inoculum is fully induced, and $\sim$0.5 if the inoculum is non-induced.
(b)~Monostability in low-repression cells exposed to 1~mM glucose
and various concentrations of extracellular TMG.}
\end{figure}

Although the experiments done by Cohn and coworkers provided clear
evidence of bistability during growth on TMG/succinate, they did not
investigate the enzyme levels at a wide variety of conditions. Recently,
Ozbudak et al measured the steady state enzyme levels at various concentrations
of TMG~\citep{ozbudak04}. To this end, they inserted into the chromosome
of \emph{Escherichia coli} MG 1655 a reporter \emph{lac} operon, i.e.,
an operon under the control of the \emph{lac} promoter, which codes
for the green fluorescent protein (GFP) instead of the \emph{lac}
enzymes. They then exposed non-induced and induced cells to a fixed
concentration of succinate, and various concentrations of TMG. It
was observed that:

\begin{enumerate}
\item When the cells are grown in the presence of succinate and various
concentrations of TMG, they exhibit bistability (Fig.~\ref{f:Oudenaarden}a).
This bistability persists even if glucose is added to the mixture
of succinate and TMG.
\item The bistability disappears if the concentration of the \emph{lac}
repressor is reduced $\sim$40-fold by transfecting the cells with
the \emph{lac} operator (Fig.~\ref{f:Oudenaarden}b).
\end{enumerate}
They also showed that these two observations were mirrored by the
bifurcation diagram for the Chung-Stephanopoulos model.

The Yagil \& Yagil model of \emph{lac} induction is not consistent
with the structure of the \emph{lac} operon and repressor. The \emph{lac}
operon contains two auxiliary operators, $O_{2}$ and $O_{3}$, in
addition to the main operator, $O_{1}$, and the \emph{lac} repressor
contains four inducer-binding sites~\citep{Lewis2005}. Furthermore,
these structural features play a crucial role in the formation of
DNA loops, the key determinants of \emph{lac} repression~\citep{Oehler1990,Oehler1994}
and induction~\citep{Oehler2006}. Molecular models taking due account
of the 3 operators and 4 inducer-binding sites yield the \emph{lac}
induction rate\begin{equation}
r_{e}^{+}\equiv V_{e}\frac{1}{1+\alpha/\left(1+K_{x}x\right)^{2}+\hat{\alpha}/\left(1+K_{x}x\right)^{4}},\label{eq:NewKinetics}\end{equation}
where $K_{x}$ is the association constant for repressor-inducer binding,
and $\alpha,\hat{\alpha}$ are related to the \emph{lac} repression
stemming from repressor-operator binding and DNA looping, respectively~\citep{Kuhlman2007,Narang2007b,Santillan2007}.
In wild-type \emph{lac}, the repression, $1+\alpha+\hat{\alpha}$,
is 1300, and the bulk of this repression is due to DNA looping ($\alpha\approx20$,
$\hat{\alpha}\approx1250$) mediated by the interaction of repressor-bound
$O_{1}$ with $O_{2}$ and $O_{3}$~\citep{Oehler1990,Oehler1994}.
The first goal of this work is to determine if the dynamics of the
Chung-Stephanopoulos model are significantly altered by these more
realistic kinetics. To this end, we consider the modified Chung-Stephanopoulos
model in which the induction rate is replaced by eq.~(\ref{eq:NewKinetics}).
We show that the dynamics of this modified model are in quantitative
agreement with the data.

Ozbudak et al also studied the growth on lactose and lactose + glucose~\citep[p.~2 of Supplement]{ozbudak04}.
They found that when non-induced cells (pregrown on succinate) are
exposed to various concentrations of lactose and lactose + glucose,
the green fluorescence of the cells has a unimodal distribution after
4~hours of growth. They did not report any experiments with induced
cells. However, the data for TMG/succinate shows that the green fluorescence
of non-induced cells has a bimodal distribution near the upper limit
of the bistable region (corresponding to extracellular TMG levels
of 15--30~$\mu$M in Fig.~\ref{f:Oudenaarden}a). The absence of
such a bimodal distribution led them to conclude that bistability
does not occur during growth on lactose and lactose + glucose. This
is consistent with the data obtained by Benzer, but contradicts the
data shown in Fig.~\ref{f:CohnLoomis}b. The second goal of this
work is to seek an explanation for the absence of bistability during
growth on lactose, and the conflicting results for lactose + glucose.

These experimental results have spurred the development of several
mathematical models, most of which are concerned with the disappearance
of bistability during growth on lactose. Thus far, two mechanisms
have been proposed.

The first mechanism proposes that during growth on lactose, the induced
cells outgrow the non-induced cells. It seems unlikely that this mechanism,
by itself, can explain the data. To see this, suppose that the non-induced
inoculum used in the experiments contains 10\% induced cells, which
double every hour. As a worst-case scenario, assume that the non-induced
cells do not grow at all. Then, after 4~h, almost 40\% of the population
is still non-induced, which is far from the unimodal distribution
observed in the experiments.

\begin{figure}[t]
\noindent \begin{centering}\subfigure[]{\includegraphics[width=2.5in,keepaspectratio]{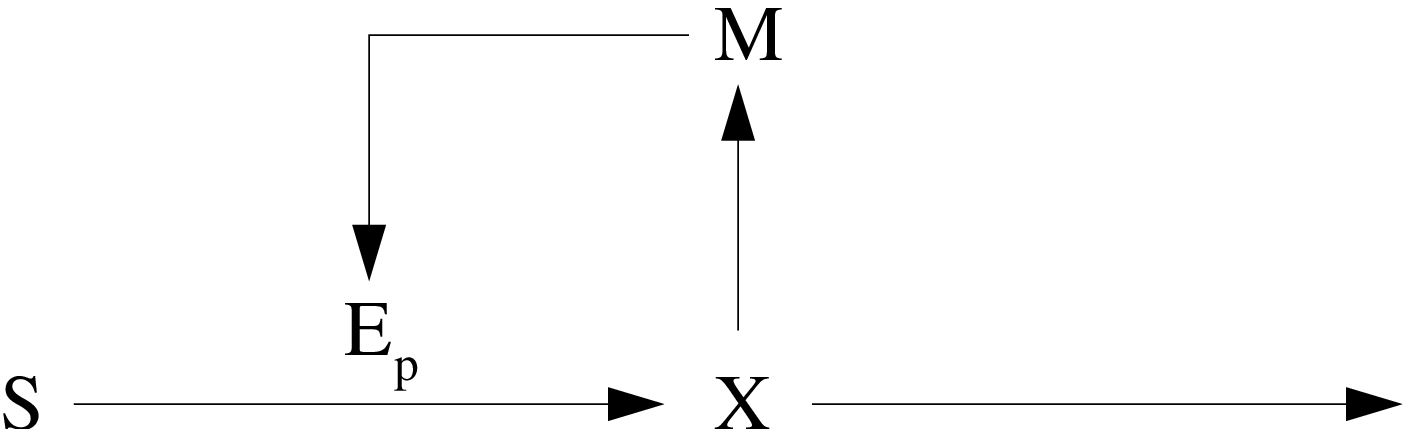}}\hspace*{0.5in}\subfigure[]{\includegraphics[width=2.5in,keepaspectratio]{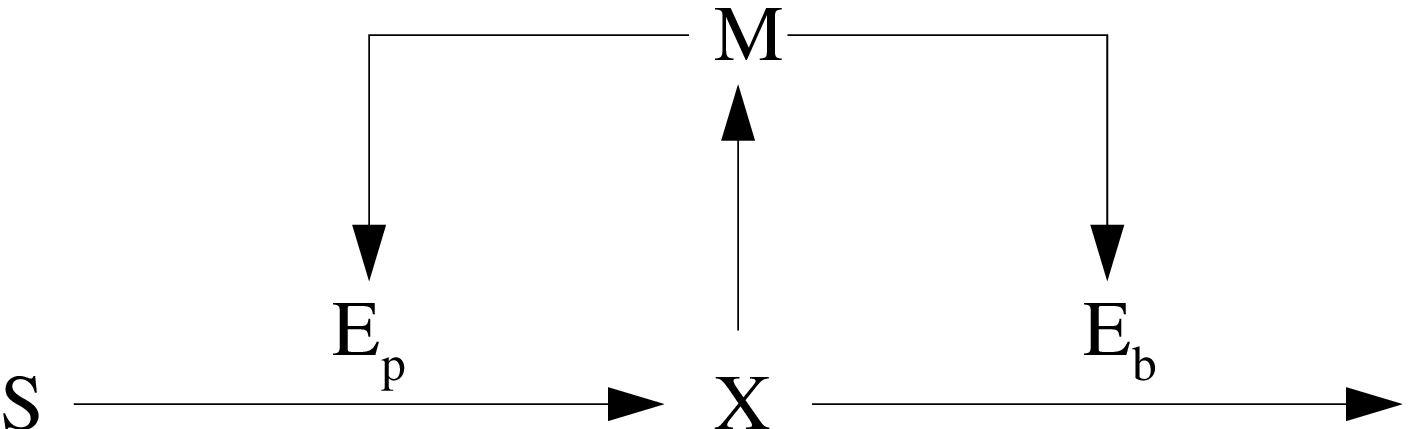}}\par\end{centering}

\caption{\label{f:Savageau}The dynamics of the inducer are different during
growth on (a)~TMG/succinate and (b)~lactose (adapted from~\citealp{Savageau2001}).
Here, $S$ denotes extracellular TMG or lactose; $X$ denotes intracellular
TMG or lactose; $M$ denotes \emph{lac} mRNA; and $E_{p},E_{b}$ denote
\emph{lac} permease and $\beta$-galactosidase, respectively. (a)~Intracellular
TMG is discharged from the cell by inducer expulsion, a process that
is independent of \emph{lac} enzymes. (b)~Intracellular lactose is
metabolised by $\beta$-galactosidase ($E_{b}$).}
\end{figure}

The second mechanism proposes that bistability does not even exist
because positive feedback is suppressed during growth on lactose or
lactose + glucose~\citep{Hoek2006,Mahaffy1999,Santillan2007,Savageau2001}.
More precisely, during growth on TMG/succinate, bistability is feasible
because of strong positive feedback: The permease stimulates the accumulation
of intracellular TMG, which in turn promotes the synthesis of even
more permease (Fig.~\ref{f:Savageau}a). The destabilizing effect
of this positive feedback produces bistability by overcoming the stabilizing
effect of dilution. During growth on lactose, the positive feedback
is suppressed because hydrolysis of lactose by $\beta$-galactosidase
attenuates the stimulatory effect of the permease (Fig.~\ref{f:Savageau}b).
It is attenuated even more during growth on lactose + glucose because
in the presence of glucose, enzyme IIA$^{\textnormal{glc}}$ is dephosphorylated,
and inhibits the permease by binding to it~\citep{Santillan2007}.

In all these models, the differences in the dynamics of growth on
TMG/succinate, lactose, and lactose + glucose are attributed entirely
to changes in the destabilizing effect of positive feedback generated
by induction. Here, we show that the stabilizing effect of dilution
also changes dramatically with the medium composition, and this has
equally profound effects on the dynamics of the \emph{lac} operon.
Specifically:

\begin{enumerate}
\item The stabilizing effect of dilution is much stronger during growth
on lactose (as opposed to growth on TMG/succinate). Indeed, during
growth on TMG/succinate, the dilution rate of the \emph{lac} enzymes
is proportional to their level, $e$, because the specific growth
rate does not depend on the activity of these enzymes --- it is completely
determined by the concentration of succinate. However, during growth
on lactose, the dilution rate is proportional to $e^{2}$ because
the specific growth rate is proportional to the lactose uptake rate,
which, in turn, is proportional to the activity of \emph{lac} permease.
We show that this stronger stabilizing effect of dilution suppresses
bistability on lactose even in the presence of the intense positive
feedback.
\item The enhanced stabilizing effect of dilution is attenuated once again
during growth on lactose + glucose. This is because the specific growth
rate on glucose, a non-galactosidic carbon source like succinate,
is independent of $e$. Thus, in the presence of glucose, the dilution
rate once again contains a term that is proportional to $e$, and
the dynamics become similar to those on TMG/succinate, i.e., bistability
is feasible, provided the positive feedback is sufficiently large.
\end{enumerate}
These results imply that the \emph{lac} operon is much more susceptible
to bistability in the presence of non-galactosidic carbon sources,
since they serve to suppress the stabilizing effect of dilution.

\section{The model}

\begin{figure}[t]
\noindent \begin{centering}\includegraphics[width=3.5in,keepaspectratio]{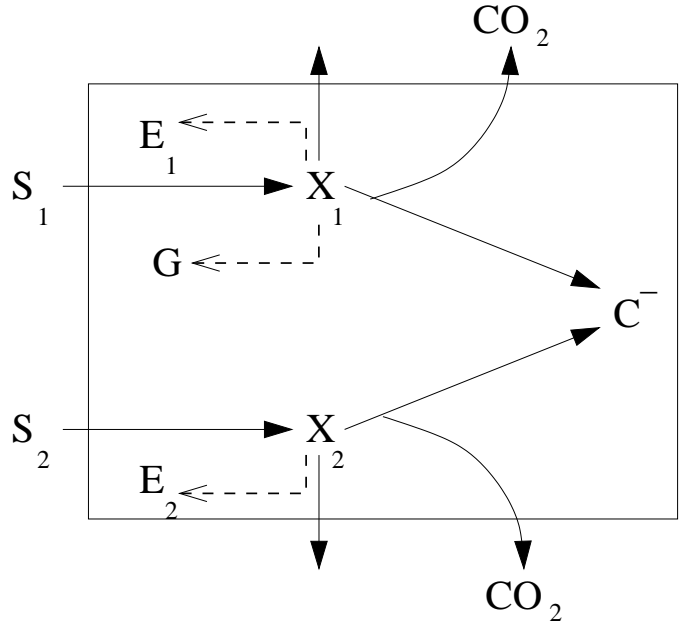}\par\end{centering}

\caption{\label{f:Scheme}Kinetic scheme of the model. }
\end{figure}

\begin{figure}[t]
\noindent \begin{centering}\includegraphics[width=3in,height=2in]{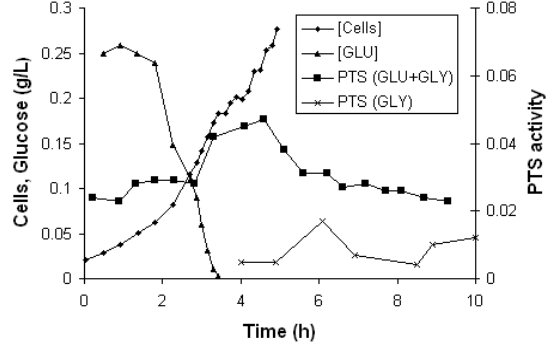}\par\end{centering}

\caption{\label{f:PTSdata}The phosphotransferase system of enzymes is inducible~\citep{Bettenbrock2006}.
During batch growth of \emph{E. coli} on glycerol, the PTS activity
($\times$) is $\sim0.01$. The concentrations of cells ($\blacklozenge$)
and glucose ($\blacktriangle$) correspond to the growth on glucose
+ glycerol. During the first phase of diauxic growth on glucose +
glycerol, the PTS activity ($\blacksquare$) increases to $\sim$0.05. }
\end{figure}

Fig.~\ref{f:Scheme} shows the kinetic scheme of the model. Here,
$S_{1},S_{2}$ denote the exogenous lactose (or TMG) and glucose,
respectively; $E_{1},E_{2}$ denote the transport enzymes for lactose
(or TMG) and glucose, respectively; $X_{1},X_{2}$ denote internalized
lactose (or TMG) and glucose, respectively; $G$ denotes the GFP synthesized
by the reporter \emph{lac} operon; and $C^{-}$ denotes all intracellular
components except $E_{i}$, $X_{i}$, and $G$ (thus, it includes
precursors, free amino acids, and macromolecules). We assume that:

\begin{enumerate}
\item The concentrations of the intracellular components, denoted $e_{i}$,
$x_{i}$, $g$, and $c^{-}$, are based on the dry weight of the cells
(g per g dry weight of cells, i.e., g~gdw$^{-1}$). The concentrations
of the exogenous substrate and cells, denoted $s_{i}$ and $c$, are
based on the volume of the reactor (g/L and gdw/L, respectively).
The rates of all the processes are based on the dry weight of the
cells (g~gdw$^{-1}$~h$^{-1}$). We shall use the term \emph{specific
rate} to emphasize this point.\\
The choice of these units implies that if the concentration of any
intracellular component, $Z$, is $z$ g~gdw$^{-1}$, then the evolution
of $z$ in batch cultures is given by\[
\frac{dz}{dt}=r_{z}^{+}-r_{z}^{-}-\left(\frac{1}{c}\frac{dc}{dt}\right)z\]
where $r_{z}^{+}$ and $r_{z}^{-}$ denote the specific rates of synthesis
and degradation of $Z$ in g~gdw$^{-1}$~h$^{-1}$.
\item The specific uptake rate of $S_{i}$, denoted $r_{s,i}$, follows
the modified Michaelis-Menten kinetics, $r_{s,i}\equiv V_{s,i}e_{i}s_{i}/(K_{s,i}+s_{i})$.
\item In the case of glucose or lactose, part of the internalized substrate,
denoted $X_{i}$, is expelled into the environment. The remainder
is converted to $C^{-}$, and oxidized to ${\rm CO_{2}}$ for generating
energy.

\begin{enumerate}
\item The specific rate of expulsion of $X_{i}$ follows first-order kinetics,
i.e., $r_{x,i}^{-}\equiv k_{x,i}^{-}x_{i}$.
\item The conversion of $X_{i}$ to $C^{-}$ and ${\rm CO_{2}}$ also follows
first-order kinetics, i.e., $r_{x,i}^{+}\equiv k_{x,i}^{+}x_{i}$.
\item The specific rate of synthesis of $C^{-}$ from $X_{i}$ is $Y_{i}r_{x,i}^{+}$,
where $Y_{i}$ is a constant (which will turn out later to be effectively
equal to the yield of biomass on $S_{i}$).
\end{enumerate}
\item Internalized TMG is completely expelled --- it does not support biosynthesis
or respiration.
\item The internalized substrates induce the synthesis of the enzymes and
GFP.

\begin{enumerate}
\item The specific synthesis rate of the lactose enzymes, $E_{1}$, follows
the kinetics\begin{equation}
r_{e,1}\equiv V_{e,1}\frac{1}{1+\alpha_{1}/\left(1+K_{x,1}x_{1}\right)^{2}+\hat{\alpha}_{1}/\left(1+K_{x,1}x_{1}\right)^{4}}\label{eq:Induction}\end{equation}
where $K_{x,1}$ is the association constant for repressor-inducer
binding, and $\alpha_{1},\hat{\alpha}_{1}$ characterize the transcriptional
repression in the absence of inducer due to repressor-operator binding
and DNA looping, respectively. Both $\alpha_{1}$ and $\hat{\alpha}_{1}$
are proportional to the intracellular repressor level~\citep{Narang2007b}.\\
In wild-type cells, $\alpha_{1}=20,\hat{\alpha}_{1}=1250$~\citep[Fig.~2]{Oehler1990}.
In cells transfected with the \emph{lac} operator, the repressor levels
decrease 43-fold~\citep{ozbudak04}; hence, $\alpha_{1}\approx0,\hat{\alpha}_{1}\approx30$.
\item The specific synthesis rate of GFP follows the same kinetics as the
\emph{lac} operon, i.e.,\[
r_{GFP}\equiv V_{GFP}\frac{1}{1+\alpha_{G}/\left(1+K_{x,1}x_{1}\right)^{2}+\hat{\alpha}_{G}/\left(1+K_{x,1}x_{1}\right)^{4}}.\]
We shall assume that the promoters of the reporter and native \emph{lac}
operons are identical, so that $V_{GFP}=V_{e,1}$ and $\alpha_{G}=\alpha_{1}=20$.
However, $\hat{\alpha}_{G}<\hat{\alpha}_{1}$ because the reporter
\emph{lac} operon lacks the auxiliary operator, $O_{2}$, which precludes
the formation of DNA loops due to interaction between $O_{1}$ and
$O_{2}$. Ozbudak et al found the repression of the \emph{lac} reporter,
$1+\alpha_{G}+\hat{\alpha}_{G}$ to be 170~\citep{ozbudak04}, which
implies that $\hat{\alpha}_{G}\approx150$. If the cells are transfected
with the \emph{lac} operator, $\alpha_{G}\approx0$, $\hat{\alpha}_{G}=4$.
\item The induction of the glucose enzymes, $E_{2}$, is thought to occur
by a mechanism similar to the one that induces the \emph{lac} operon~\citep{plumbridge03}.
Specifically, in the absence of glucose, transcription of the \emph{ptsG}
operon is blocked because the repressor (Mlc) is bound to the operator.
In the presence of glucose, the enzyme II$^{\textnormal{glc}}$ sequesters
Mlc from the operator by an unknown mechanism, thus liberating the
operon for transcription. We assume that the specific synthesis rate
of $E_{2}$ has the form\[
r_{e,2}\equiv V_{e,2}\frac{1}{1+\alpha_{2}/\left(1+K_{x,2}x_{2}\right)},\]
where $K_{x,2}$ and $\alpha_{2}$ are phenomenological parameters,
i.e., they cannot be expressed in terms of parameters characterizing
the molecular interactions.\\
The data shows that the PTS level in cells growing exponentially on
glucose is roughly 5 times the level observed in cells growing exponentially
on glycerol (Fig.~\ref{f:PTSdata}). This implies that $\alpha_{2}\approx4$.
\item The synthesis of the enzymes and GFP occurs at the expense of the
biosynthetic constituents, $C^{-}$.
\item Enzyme and GFP degradation are negligibly small.
\item Non-specific diffusion of the substrates into the cell is negligibly
small. This is valid for lactose and glucose at the concentrations
typically used in the experiments. It is valid for gratuitous inducers,
such as TMG, only if the extracellular concentration is <50~$\mu$M~\citep[Fig.~4]{Herzenberg1959}.
\end{enumerate}
\end{enumerate}
In what follows, we begin by deriving the equations for growth on
TMG plus non-galactosidic substrates. We then derive the equations
for growth on lactose + glucose, from which the equations for growth
on lactose are obtained by letting the concentration of glucose be
zero.

\subsection{Growth in the presence of TMG and non-galactosidic carbon sources}

During growth in the presence of TMG and non-galactosidic carbon sources,
the mass balances yield\begin{align*}
\frac{ds_{1}}{dt} & =-\left[V_{s,1}e_{1}\frac{s_{1}}{K_{s,1}+s_{1}}-k_{x,1}^{-}x_{1}\right]c,\\
\frac{dx_{1}}{dt} & =V_{s,1}e_{1}\frac{s_{1}}{K_{s,1}+s_{1}}-k_{x,1}^{-}x_{1}-\left(\frac{1}{c}\frac{dc}{dt}\right)x_{1},\\
\frac{de_{1}}{dt} & =V_{e,1}\frac{1}{1+\alpha_{1}/\left(1+K_{x,1}x_{1}\right)^{2}+\hat{\alpha}_{1}/\left(1+K_{x,1}x_{1}\right)^{4}}-\left(\frac{1}{c}\frac{dc}{dt}\right)e_{1},\\
\frac{dg}{dt} & =V_{e,1}\frac{1}{1+\alpha_{1}/\left(1+K_{x,1}x_{1}\right)^{2}+\hat{\alpha}_{G}/\left(1+K_{x,1}x_{1}\right)^{4}}-\left(\frac{1}{c}\frac{dc}{dt}\right)g,\\
\frac{dc}{dt} & =r_{g}c\end{align*}
where $r_{g}$, the exponential growth rate on the non-galactosidic
carbon sources, is a fixed parameter (independent of the model variables,
$s_{1},x_{1},e_{1},g,c$). It is completely determined by the concentration
of the non-galactosidic carbon sources(s).

It turns out that the dynamics of the experiments can be described
by only 2 differential equations. Indeed, since the cell density remains
vanishingly small throughout the experiment, there is almost no depletion
of extracellular TMG, and $s_{1}$ remains essentially equal to its
initial value, $s_{1,0}$. Moreover, since $k_{x,1}^{-}\sim1$~min$^{-1}$~\citep[Table~1]{Kepes1960}
and $r_{g}\sim0.5$~hr$^{-1}$, $x_{1}$ attains quasisteady state
on a time scale of minutes. It follows that the dynamics of $E_{1}$
after the first few minutes are well-approximated by the equations\begin{align}
\frac{de_{1}}{dt} & =V_{e,1}\frac{1}{1+\alpha_{1}/\left(1+K_{x,1}x_{1}\right)^{2}+\hat{\alpha}_{1}/\left(1+K_{x,1}x_{1}\right)^{4}}-r_{g}e_{1},\label{eq:TMGEorig}\\
\frac{dg}{dt} & =V_{e,1}\frac{1}{1+\alpha_{1}/\left(1+K_{x,1}x_{1}\right)^{2}+\hat{\alpha}_{G}/\left(1+K_{x,1}x_{1}\right)^{4}}-r_{g}g,\label{eq:TMGgOrig}\\
x_{1} & \approx\frac{V_{s,1}}{k_{x,1}^{-}}e_{1}\sigma_{1},\;\sigma_{1}\equiv\frac{s_{1,0}}{K_{s,1}+s_{1,0}},\label{eq:TMGxorig}\end{align}
where (\ref{eq:TMGxorig}) shows the quasisteady state concentration
of intracellular TMG.

Eqs.~(\ref{eq:TMGEorig})--(\ref{eq:TMGgOrig}) imply that the maximum
steady state level of $E_{1}$ and $G$ is $V_{e,1}/r_{g}$. It is
therefore natural to define the dimensionless variables\[
\epsilon_{1}\equiv\frac{e_{1}}{V_{e,1}/r_{g}},\;\gamma\equiv\frac{g}{V_{e,1}/r_{g}},\;\chi_{1}\equiv K_{x,1}x_{1},\;\tau\equiv r_{g}t.\]
This yields the dimensionless equations\begin{align}
\frac{d\epsilon_{1}}{d\tau} & =\frac{1}{1+\alpha_{1}/\left(1+\chi_{1}\right)^{2}+\hat{\alpha}_{1}/\left(1+\chi_{1}\right)^{4}}-\epsilon_{1},\label{eq:TMGe}\\
\frac{d\gamma}{d\tau} & =\frac{1}{1+\alpha_{1}/\left(1+\chi_{1}\right)^{2}+\hat{\alpha}_{G}/\left(1+\chi_{1}\right)^{4}}-\gamma,\label{eq:TMGg}\\
\chi_{1} & =\bar{\delta}_{1}\epsilon_{1}\label{eq:TMGx}\end{align}
where \begin{equation}
\bar{\delta}_{1}\equiv\bar{\delta}_{1,m}\sigma_{1},\;\bar{\delta}_{1,m}\equiv\frac{V_{s,1}\left(V_{e,1}/r_{g}\right)/k_{x,1}^{-}}{K_{x,1}^{-1}}.\label{eq:TMGDelta1}\end{equation}
It follows from (\ref{eq:TMGx})--(\ref{eq:TMGDelta1}) that $\bar{\delta}_{1,m}$
is a measure of the quasisteady state intracellular TMG concentration
at saturating levels of extracellular TMG (measured in units of $K_{x,1}^{-1}$,
the dissociation constant for repressor-inducer binding). We can also
view $\delta_{1,m}$ as a measure of the strength of the positive
feedback: Given any $\sigma_{1}$, the strength of the positive feedback,
$\partial r_{e,1}/\partial\epsilon_{1}$, is an increasing function
of $\delta_{1,m}$. The parameter, $\bar{\delta}_{1}$, is a measure
of the intracellular TMG concentration at any given extracellular
TMG level. For a given cell type and non-galactosidic carbon source,
$\bar{\delta}_{1}$ is proportional to $\sigma_{1}$, and hence, can
be treated as a surrogate for the extracellular TMG concentration.

\subsection{Growth on lactose or lactose + glucose}

During growth in the presence of lactose and glucose, the mass balances
yield\begin{align}
\frac{ds_{i}}{dt} & =-\left(V_{s,i}e_{i}\frac{s_{i}}{K_{s,i}+s_{i}}-k_{x}^{-}x_{i}\right)c,\label{eq:MSsOrig}\\
\frac{dx_{i}}{dt} & =V_{s,i}e_{i}\frac{s_{i}}{K_{s,i}+s_{i}}-k_{x,i}^{-}x_{i}-k_{x,i}^{+}x_{i}-\left(\frac{1}{c}\frac{dc}{dt}\right)x_{i},\label{eq:MSxOrig}\\
\frac{de_{1}}{dt} & =V_{e,1}\frac{1}{1+\alpha_{1}/\left(1+K_{x,1}x_{1}\right)^{2}+\hat{\alpha}_{1}/\left(1+K_{x,1}x_{1}\right)^{4}}-\left(\frac{1}{c}\frac{dc}{dt}\right)e_{1},\label{eq:MSe1Orig}\\
\frac{dg}{dt} & =V_{e,1}\frac{1}{1+\alpha_{1}/\left(1+K_{x,1}x_{1}\right)^{2}+\hat{\alpha}_{G}/\left(1+K_{x,1}x_{1}\right)^{4}}-\left(\frac{1}{c}\frac{dc}{dt}\right)g,\label{eq:MSgOrig}\\
\frac{de_{2}}{dt} & =V_{e,2}\frac{1}{1+\alpha_{2}/\left(1+K_{x,2}x_{2}\right)}-\left(\frac{1}{c}\frac{dc}{dt}\right)e_{2},\label{eq:MSe2Orig}\\
\frac{dc^{-}}{dt} & =\sum_{i=1}^{2}\left(Y_{i}k_{x,i}^{+}x_{i}\right)-r_{e,1}-r_{GFP}-r_{e,2}-\left(\frac{1}{c}\frac{dc}{dt}\right)c^{-}\label{eq:MScMinusOrig}\end{align}
It is shown in Appendix~\ref{a:GlucoseLactoseEqns} that under the
experimental conditions, the dynamics of the enzymes and GFP are well-approximated
by the equations\begin{align}
\frac{de_{1}}{dt} & =V_{e,1}\frac{1}{1+\alpha_{1}/\left(1+K_{x,1}x_{1}\right)^{2}+\hat{\alpha}_{1}/\left(1+K_{x,1}x_{1}\right)^{4}}-\left(\phi_{1}Y_{1}V_{s,1}\sigma_{1}e_{1}+\phi_{2}Y_{2}V_{s,2}\sigma_{2}e_{2}\right)e_{1},\label{eq:MSe1Orig1}\\
\frac{dg}{dt} & =V_{e,1}\frac{1}{1+\alpha_{G}/\left(1+K_{x,1}x_{1}\right)^{2}+\hat{\alpha}_{G}/\left(1+K_{x,1}x_{1}\right)^{4}}-\left(\phi_{1}Y_{1}V_{s,1}\sigma_{1}e_{1}+\phi_{2}Y_{2}V_{s,2}\sigma_{2}e_{2}\right)g,\label{eq:MSgOrig1}\\
\frac{de_{2}}{dt} & =V_{e,2}\frac{1}{1+\alpha_{2}/\left(1+K_{x,2}x_{2}\right)}-\left(\phi_{1}Y_{1}V_{s,1}\sigma_{1}e_{1}+\phi_{2}Y_{2}V_{s,2}\sigma_{2}e_{2}\right)e_{2},\label{eq:MSe2Orig1}\\
x_{i} & =\frac{V_{s,i}}{k_{x,i}^{+}+k_{x,i}^{-}}e_{i}\sigma_{i},\label{eq:MSxOrig1}\end{align}
where \[
\sigma_{i}\equiv\frac{s_{i,0}}{K_{s,i}+s_{i,0}},\;\phi_{i}\equiv\frac{k_{x,i}^{+}}{k_{x,i}^{-}+k_{x,i}^{+}}.\]
The parameter, $\phi_{i}$, is the fraction of substrate intake that
is channeled into growth and respiration (the remainder is expelled
into the medium). The second term in eqs.~(\ref{eq:MSe1Orig1})--(\ref{eq:MSe2Orig1})
represents the dilution rate of $E_{1}$, $G$, and $E_{2}$, respectively.%
\footnote{These equations are formally similar to the model considered in~\citealp{Narang2007a},
the main difference being that the induction kinetics were assumed
to follow Yagil \& Yagil kinetics.%
}

It follows from (\ref{eq:MSe1Orig1})--(\ref{eq:MSe2Orig1}) that
during single-substrate growth on $S_{i}$, the steady state activity
of $E_{i}$ is at most\[
\sqrt{\frac{V_{e,i}}{\phi_{i}Y_{i}V_{s,i}\sigma_{i}}},\]
and the maximum specific growth rate is at most\[
\phi_{i}Y_{i}V_{s,i}\sqrt{\frac{V_{e,i}}{\phi_{i}Y_{i}V_{s,i}\sigma_{i}}}\sigma_{i}=\sqrt{\phi_{i}Y_{i}V_{s,i}V_{e,i}\sigma_{i}}.\]
Thus, we are led to define the dimensionless variables\[
\epsilon_{i}\equiv\frac{e_{i}}{\sqrt{V_{e,i}/\left(\phi_{i}Y_{i}V_{s,i}\sigma_{i}\right)}},\gamma\equiv\frac{g}{\sqrt{V_{e,1}/\left(\phi_{1}Y_{1}V_{s,1}\sigma_{1}\right)}},\;\chi_{i}\equiv\; K_{x,i}x_{i},\;\tau\equiv t\sqrt{\phi_{1}Y_{1}V_{s,1}V_{e,1}\sigma_{1}},\]
which yield the dimensionless equations\begin{align}
\frac{d\epsilon_{1}}{d\tau} & =\frac{1}{1+\alpha_{1}/\left(1+\chi_{1}\right)^{2}+\hat{\alpha}_{1}/\left(1+\chi_{1}\right)^{4}}-\left(\epsilon_{1}+\alpha\epsilon_{2}\right)\epsilon_{1},\label{eq:MSe1}\\
\frac{d\gamma}{d\tau} & =\frac{1}{1+\alpha_{1}/\left(1+\chi_{1}\right)^{2}+\hat{\alpha}_{G}/\left(1+\chi_{1}\right)^{4}}-\left(\epsilon_{1}+\alpha\epsilon_{2}\right)\gamma,\label{eq:MSg}\\
\frac{d\epsilon_{2}}{d\tau} & =\alpha\frac{1}{1+\alpha_{2}/\left(1+\chi_{2}\right)}-\left(\epsilon_{1}+\alpha\epsilon_{2}\right)\epsilon_{2},\label{eq:MSe2}\\
\chi_{i} & =\delta_{i}\epsilon_{i},\; i=1,2\label{eq:MSx}\end{align}
with dimensionless parameters\begin{align}
\alpha & \equiv\frac{\sqrt{\phi_{2}Y_{2}V_{s,2}V_{e,2}\sigma_{2}}}{\sqrt{\phi_{1}Y_{1}V_{s,1}V_{e,1}\sigma_{1}}},\label{eq:alpha}\\
\delta_{i} & \equiv\delta_{i,m}\sqrt{\sigma_{i}},\;\delta_{i,m}\equiv\frac{K_{x,i}}{k_{x,i}^{-}+k_{x,i}^{+}}\sqrt{\frac{V_{s,i}V_{e,i}}{\phi_{i}Y_{i}}}.\label{eq:MSdelta1}\end{align}
Here, $\alpha$, is a measure of the specific growth rate on $S_{2}$
relative to that on $S_{1}$, and $\delta_{i,m}$ is a measure of
the quasisteady state concentration of $X_{i}$ at saturating concentrations
of $S_{i}$ (or equivalently, the strength of the positive feedback
generated by induction of $E_{i}$).

Unlike TMG, lactose is rapidly metabolized to support growth and respiration.
It follows that the ability of the cells to accumulate the intracellular
substrate, (and hence, the strength of the positive feedback) is smaller
during growth on lactose, i.e., $\delta_{1,m}<\bar{\delta}_{1,m}$.
Indeed, (\ref{eq:TMGDelta1}) and (\ref{eq:MSdelta1}) imply that
\[
\frac{\delta_{1,m}}{\bar{\delta}_{1,m}}=\frac{r_{g}}{\sqrt{\phi_{1}Y_{1}V_{s,1}V_{e,1}}}\frac{k_{x,1}^{-}}{k_{x,1}^{-}+k_{x,1}^{+}}.\]
Since $r_{g}$, the specific growth rate in the experiments with TMG,
is comparable to $\sqrt{\phi_{1}Y_{1}V_{s,1}V_{e,1}}$, a measure
of the maximum specific growth rate on lactose, we have\[
\frac{\delta_{1,m}}{\bar{\delta}_{1,m}}\approx\frac{k_{x,1}^{-}}{k_{x,1}^{-}+k_{x,1}^{+}},\]
which is less than 1.

In the particular case of growth on lactose, $\sigma_{2}=\alpha=0$,
and the above equations become \begin{align}
\frac{d\epsilon_{1}}{d\tau} & =\frac{1}{1+\alpha_{1}/\left(1+\chi_{1}\right)^{2}+\hat{\alpha}_{1}/\left(1+\chi_{1}\right)^{4}}-\epsilon_{1}^{2},\label{eq:LactoseE}\\
\frac{d\gamma}{d\tau} & =\frac{1}{1+\alpha_{1}/\left(1+\chi_{1}\right)^{2}+\hat{\alpha}_{G}/\left(1+\chi_{1}\right)^{4}}-\epsilon_{1}\gamma,\label{eq:LactoseG}\\
\chi_{1} & =\delta_{1}\epsilon_{1}.\label{eq:LactoseX}\end{align}
Note that (\ref{eq:LactoseE}) is formally similar to (\ref{eq:TMGe}),
the only difference being that the dilution rate is proportional to
$\epsilon_{1}^{2}$ rather than $\epsilon_{1}$. This reflects the
fact that during growth on lactose, the specific growth rate is proportional
to the activity of $E_{1}$.

\section{Results and Discussion}

We note at the outset that the steady state GFP level is completely
determined by the steady state activity of the \emph{lac} enzymes.
More precisely, eqs.~(\ref{eq:TMGe})--(\ref{eq:TMGg}) and (\ref{eq:MSe1})--(\ref{eq:MSg})
imply that\begin{equation}
\frac{\gamma}{\epsilon_{1}}=\frac{1+\alpha_{1}/\left(1+\chi_{1}\right)^{2}+\hat{\alpha}_{1}/\left(1+\chi_{1}\right)^{4}}{1+\alpha_{1}/\left(1+\chi_{1}\right)^{2}+\hat{\alpha}_{G}/\left(1+\chi_{1}\right)^{4}},.\label{eq:ratioGFPenzyme}\end{equation}
where $\chi_{1}$ is given by (\ref{eq:TMGx}) or (\ref{eq:MSx}).
In what follows, we shall focus on the variation of the steady state
enzyme activity, $\epsilon_{1}$, with the extracellular TMG ($\bar{\delta}_{1}$)
or lactose ($\delta_{1}$) concentration. Given this relation, the
steady state GFP level completely determined by (\ref{eq:ratioGFPenzyme}).

If the repression characteristics of the \emph{lac} reporter were
identical to those of native \emph{lac} ($\alpha_{1}=\alpha_{G}$,
$\hat{\alpha}_{1}=\hat{\alpha}_{G}$), $\gamma$ would be identical
to $\epsilon_{1}$. However, since $\hat{\alpha}_{G}=150$ is significantly
smaller than $\hat{\alpha}_{1}=1250$, the ratio, $\gamma/\epsilon_{1}$,
is a decreasing function of $\chi_{1}$. Now the inducer levels, $\chi_{1}$,
are vanishingly small in non-induced cells, and very large in induced
cells. Hence, (\ref{eq:ratioGFPenzyme}) implies that \[
\frac{\gamma_{\textnormal{non-induced}}}{\epsilon_{1,\textnormal{non-induced}}}\approx8,\;\frac{\gamma_{\textnormal{induced}}}{\epsilon_{1,\textnormal{induced}}}\approx1\Rightarrow\frac{\left(\gamma_{\textnormal{induced}}/\gamma_{\textnormal{non-induced}}\right)}{\left(\epsilon_{1,\textnormal{induced}}/\epsilon_{1,\textnormal{non-induced}}\right)}\approx\frac{1}{8},\]
i.e., ratio of the GFP levels in induced and non-induced cells is
significantly smaller than the corresponding ratio of the enzyme activities.
We shall appeal to this fact later.

In what follows, we consider the growth on TMG, lactose, and lactose
+ glucose. In the first two cases, we also study the dynamics in the
absence of DNA looping ($\hat{\alpha}_{1}=0$). We consider this biologically
unrealistic scenario because it yields useful intuitive insights.

\subsection{Growth on TMG and non-galactosidic carbon sources}

\subsubsection{No DNA looping}

In this case, the enzyme dynamics are given by (\ref{eq:TMGe})--(\ref{eq:TMGx})
with $\hat{\alpha}_{1}=0$, and the steady states satisfy the equation
\begin{equation}
f(\epsilon_{1})\equiv\frac{1}{1+\alpha_{1}/\left(1+\bar{\delta}_{1}\epsilon_{1}\right)^{2}}-\epsilon_{1}=0.\label{eq:Case1Ess}\end{equation}
Since the induction rate lies between $1/(1+\alpha_{1}$) and 1, so
does the steady state enzyme activity.

\begin{figure}[t]
\noindent \begin{centering}\subfigure[]{\includegraphics[width=2.5in,keepaspectratio]{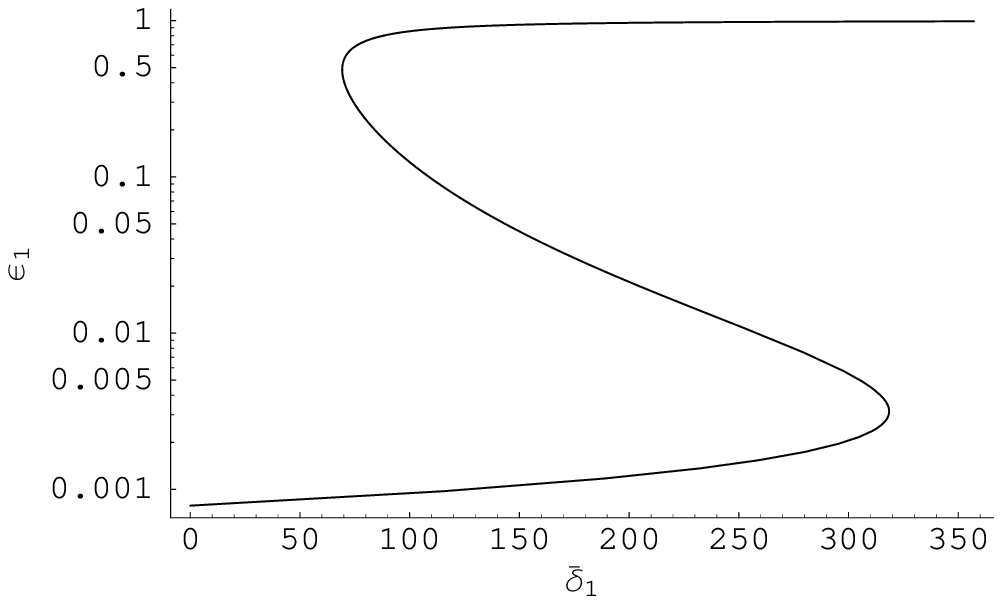}}\hspace*{0.1in}\subfigure[]{\includegraphics[width=2.5in,keepaspectratio]{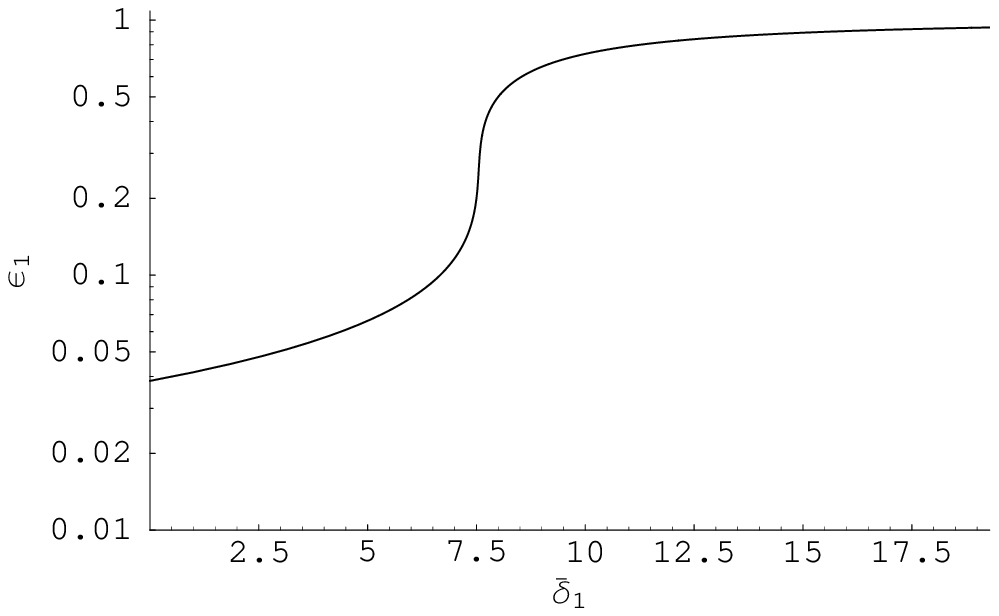}}\par\end{centering}

\caption{\label{f:Case1SS}Variation of the steady state enzyme activity ($\epsilon_{1}$)
with extracellular TMG level ($\bar{\delta}_{1}$) at high and low
repression levels. (a)~At high repression levels ($\alpha_{1}=1200$),
there is a range of extracellular TMG concentrations at which the
enzyme activity is bistable. (b)~At low repression levels ($\alpha_{1}=25$),
there is a unique enzyme activity at every extracellular TMG level.}
\end{figure}

Eq.~(\ref{eq:Case1Ess}) captures the steady state data shown in
Fig.~\ref{f:Oudenaarden}. Indeed, (\ref{eq:Case1Ess}) implies that
\[
\bar{\delta}_{1}(\epsilon_{1})\equiv\sqrt{\frac{\alpha_{1}}{\epsilon_{1}\left(1-\epsilon_{1}\right)}}-\frac{1}{\epsilon_{1}}.\]
Given any $\alpha_{1}$, the parametric curve, $\left(\bar{\delta}_{1}(\epsilon_{1}),\epsilon_{1}\right)$,
$1/(1+\alpha_{1})<\epsilon_{1}<1$, yields the variation of the steady
state enzyme activity with $\bar{\delta}_{1}$, a surrogate for the
extracellular TMG concentration. If the repression is large, the locus
of steady states is hysteretic (Fig.~\ref{f:Case1SS}a). If the repression
is small, there is a unique enzyme activity at every extracellular
TMG level (Fig.~\ref{f:Case1SS}b).

\begin{figure}[t]
\noindent \begin{centering}\subfigure[]{\includegraphics[width=3in,keepaspectratio]{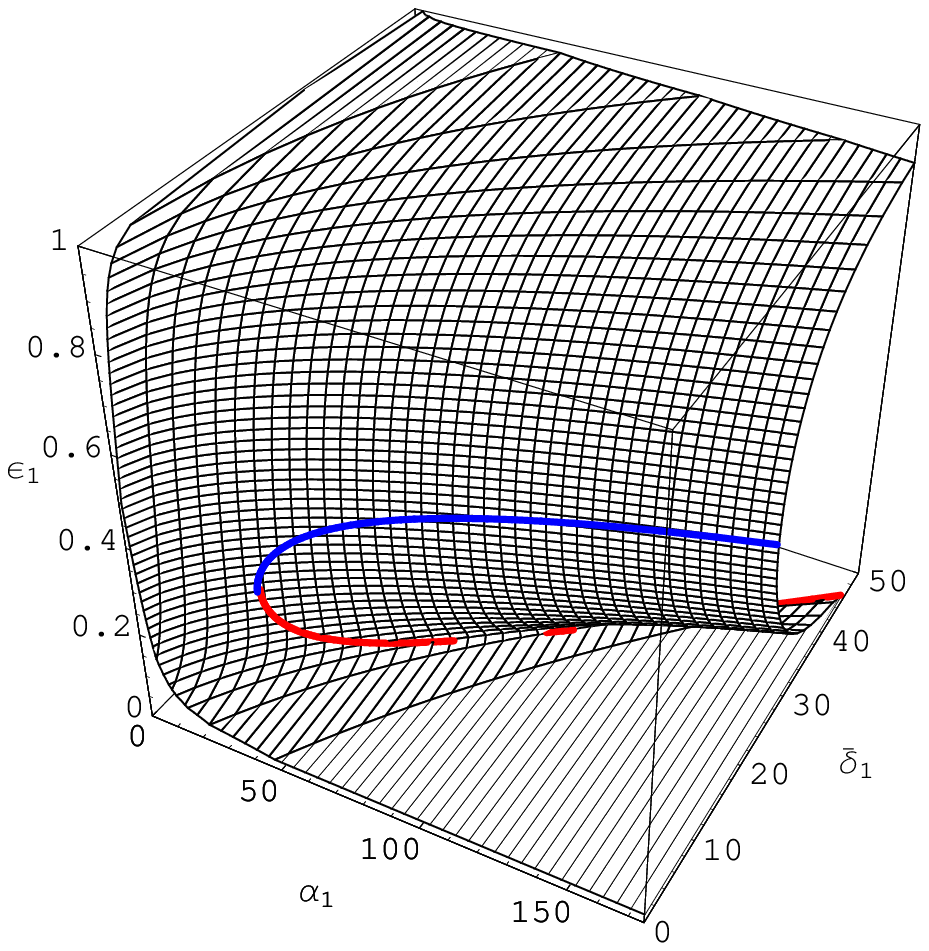}}\hspace*{0.1in}\subfigure[]{\includegraphics[width=3in,height=2.5in]{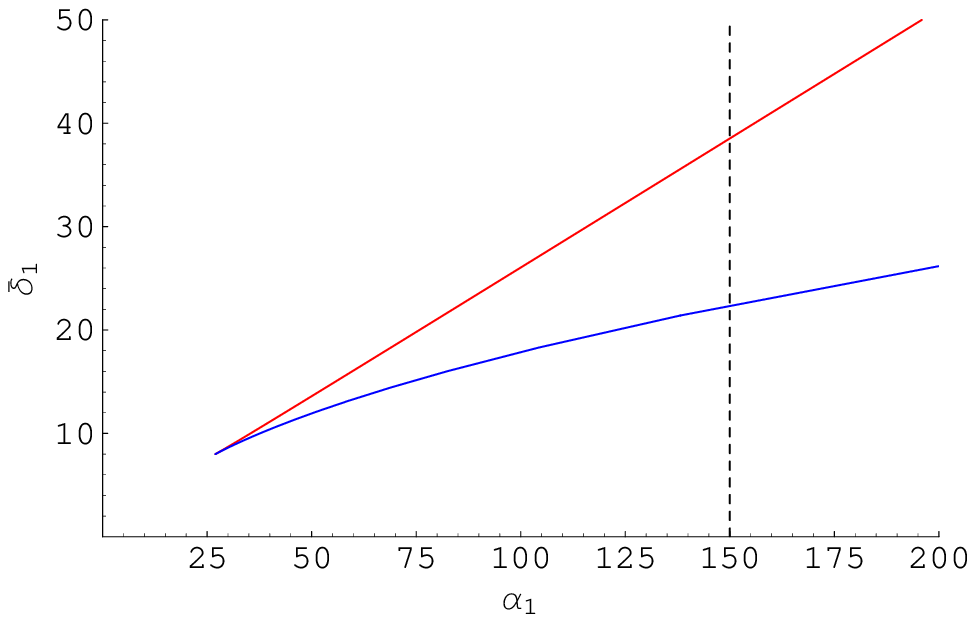}}\par\end{centering}

\caption{\label{f:TMGbdNoLooping}(a) The surface of steady state enzyme levels
for growth on TMG. The surface contains two folds, which are represented
by the red and blue curves. (b) The bifurcation diagram for growth
on TMG, obtained by projecting the folds in (a) onto the $\alpha_{1}\bar{\delta}_{1}$-plane.
Bistability occurs precisely if $\alpha_{1}$ and $\bar{\delta}_{1}$
lie in the region between the red and blue curves. The dashed line
show the path along which $\bar{\delta}_{1}$ changes when a given
cell type (fixed $\alpha_{1}$) is exposed to various extracellular
TMG concentrations.}
\end{figure}

The loss of bistability at low repression levels becomes more transparent
if we plot the surface of steady states as a function of the parameters,
$\alpha_{1},\bar{\delta}_{1}$ (Fig.~\ref{f:TMGbdNoLooping}a). The
steady states in Fig.~\ref{f:Case1SS} were obtained by varying $\bar{\delta}_{1}$
at fixed $\alpha_{1}$. These steady states are represented in Fig.~\ref{f:TMGbdNoLooping}a
by the intersection of the plane, $\alpha_{1}=\textnormal{constant}$,
with the surface of steady states. Now, at high repression levels,
the surface contains two folds (red and blue curves). Thus, the steady
states of high-repression cells correspond to a curve with two folds.
As the repression level decreases, the two folds approach each other
until they merge and disappear. The steady states of low-repression
cells therefore increase monotonically.

Ozbudak et al showed that the transition from bistability to monostability
occurs at a repression level lying between 5 and 50. We show below
that the model captures this result. To this end, we begin by deriving
the equations describing the folds of the steady state surface. Evidently,
the folds satisfy (\ref{eq:Case1Ess}) because they lie on the steady
state surface. Since the folds are singular points of the surface,
they also satisfy the equation\begin{equation}
f_{\epsilon_{1}}(\epsilon_{1})=\frac{2\alpha_{1}\bar{\delta}_{1}/\left(1+\bar{\delta}_{1}\epsilon_{1}\right)^{3}}{\left[1+\alpha_{1}/\left(1+\bar{\delta}_{1}\epsilon_{1}\right)^{2}\right]^{2}}-1=0.\label{eq:Case1EBifn}\end{equation}
It is shown in Appendix~\ref{a:TMGbd} that eqs.~(\ref{eq:Case1Ess})--(\ref{eq:Case1EBifn})
define a curve in the $\alpha_{1}\bar{\delta}_{1}\epsilon_{1}$-space
with the parametric representation\begin{equation}
\epsilon_{1}(\chi_{1})=\frac{\chi_{1}-1}{2\chi_{1}},\;\alpha_{1}(\chi_{1})=\frac{\left(1+\chi_{1}\right)^{3}}{\chi_{1}-1},\;\bar{\delta}_{1}(\chi_{1})=\frac{2\chi_{1}^{2}}{\chi_{1}-1}.\label{eq:TMGnoLooping}\end{equation}
As $\chi_{1}$ increases from $1^{+}$ to $\infty$, this parametric
representation traces the folds of the steady state surface.

The bifurcation diagram for a system refers to the classification
of its dynamics in parameter space. For the system at hand, it is
obtained by projecting the folds onto the $\alpha_{1}\bar{\delta}_{1}$-plane
(Fig.~\ref{f:TMGbdNoLooping}b). Evidently, multiple steady states
occur precisely when the parameters, $\alpha_{1}$ and $\bar{\delta}_{1}$,
lie between the blue and red curves in Fig.~\ref{f:TMGbdNoLooping}b.
These two curves meet at a cusp characterized by the conditions\begin{align*}
\frac{d\alpha_{1}}{d\chi_{1}} & =2\frac{\left(1+\chi_{1}\right)^{2}\left(\chi_{1}-2\right)}{\left(\chi_{1}-1\right)^{2}}=0,\\
\frac{d\bar{\delta}_{1}}{d\chi_{1}} & =2\frac{\chi_{1}\left(\chi_{1}-2\right)}{\left(\chi_{1}-1\right)^{2}}=0,\end{align*}
which imply that the cusp occurs at $\chi_{1}=2$, and its coordinates
are $\alpha_{1}=27$, $\bar{\delta}_{1}=8$. It follows that bistability
is feasible only if $\alpha_{1}>27$, which lies half-way between
the experimentally measured bounds (5 and 50). Thus, the model yields
results that are consistent with the data.

Fig.~\ref{f:TMGbdNoLooping}b is characterized by two properties:
(a) There is no bistability at small $\alpha_{1}$, and (b) bistability
is feasible at high $\alpha_{1}$, provided $\bar{\delta}_{1}$ lies
lies within a finite interval. To a first degree of approximation,
these properties reflect the following physical fact: Bistability
occurs precisely when the induction rate is cooperative. To see this,
observe that to a first degree of approximation, we can identify the
existence of cooperative kinetics with the existence of an inflection
point on the induction curve. Since the induction rate has an inflection
point at $\epsilon_{1}=\left(\sqrt{\alpha_{1}/3}-1\right)/\bar{\delta}_{1}$,
it follows that:

\begin{enumerate}
\item If $\alpha_{1}$ is small, there is no inflection point. Hence, the
induction rate is not cooperative, i.e., its slope decreases monotonically
with $\epsilon_{1}$, and bistability is impossible.
\item If $\alpha_{1}$ is large, the induction rate contains an inflection
point, but it is cooperative, in effect, only within a finite interval
of $\bar{\delta}_{1}$. At large $\bar{\delta}_{1}$, the inflection
point is so close to 0 that the induction rate is not cooperative,
except at vanishingly small enzyme levels. At small $\bar{\delta}_{1}$,
the inflection point is so large compared to 1 that the induction
rate is effectively linear on the interval, $0<\epsilon_{1}<1$, containing
the steady state. Thus, bistability is feasible only in a finite interval
of $\bar{\delta}_{1}$.
\end{enumerate}
In other words, the cooperativity of the quasisteady state induction
rate depends on the intracellular parameter, $\alpha_{1}$, and the
state of the environment ($\bar{\delta}_{1}$ is proportional to $\sigma_{1}$).
If $\alpha_{1}$ is small, bistability is impossible because there
is no extracellular TMG level that can make the induction rate cooperative.
If $\alpha_{1}$ is large, there is a finite range of extracellular
TMG levels at which the induction rate is cooperative, and bistability
is feasible.

\subsubsection{DNA looping}

We assumed above that the repression was entirely due to repressor-operator
binding. In reality, the repression is dominated by DNA looping. Since
the induction kinetics are qualitatively different in the presence
of DNA looping, it is relevant to ask if the foregoing conclusions
are dramatically altered when $\hat{\alpha}_{1}>0$. We show below
that the bifurcation diagram is essentially unchanged if $0<\hat{\alpha}_{1}<16$,
but it is qualitatively different at larger values of $\hat{\alpha}_{1}$.

\begin{figure}[t]
\noindent \begin{centering}\subfigure[]{\includegraphics[width=3in,keepaspectratio]{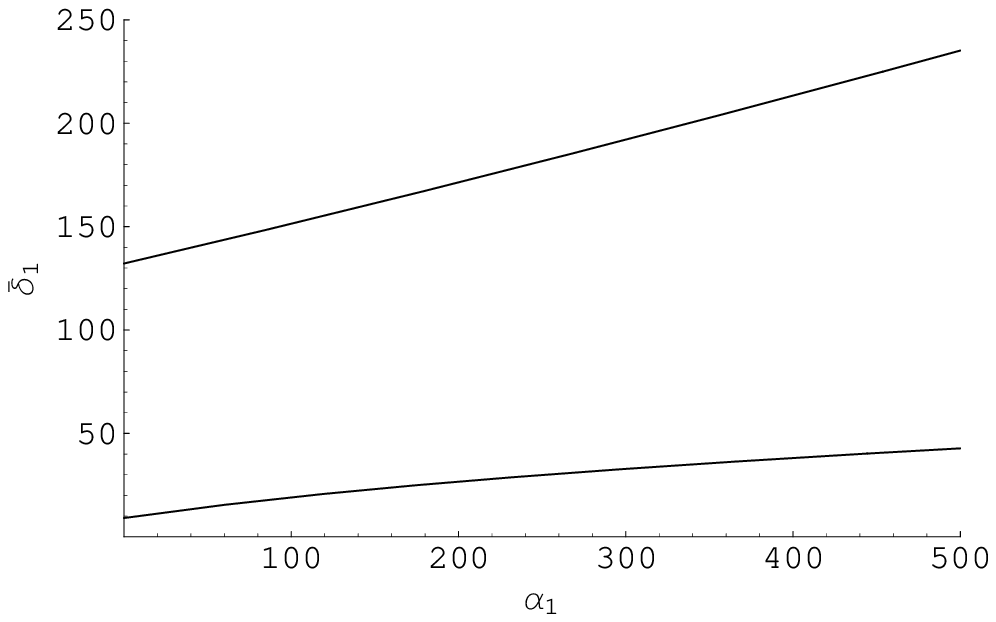}}\hspace*{0.1in}\subfigure[]{\includegraphics[width=3in,keepaspectratio]{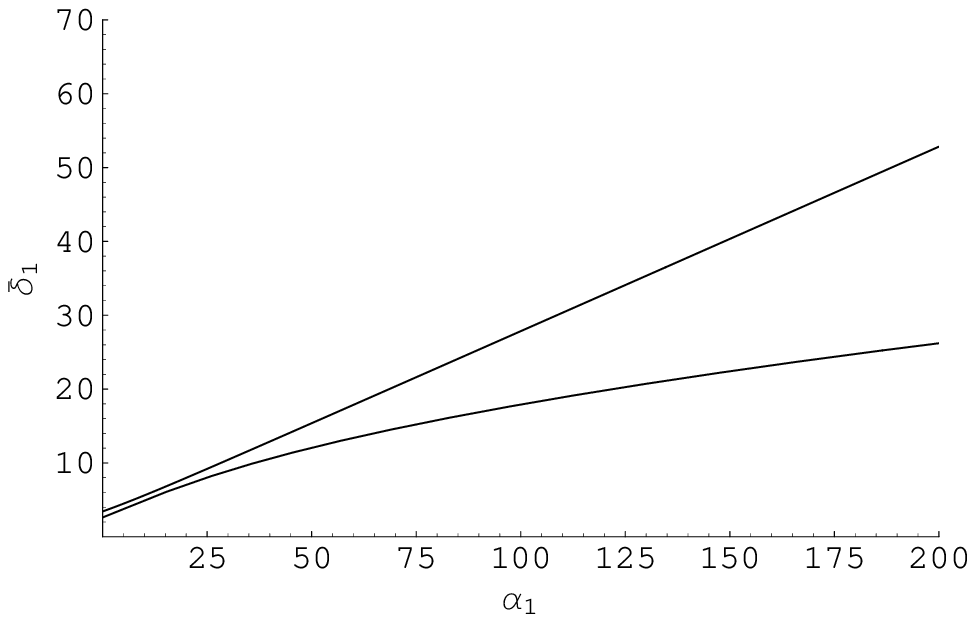}}\par\end{centering}

\noindent \begin{centering}\subfigure[]{\includegraphics[width=3in,keepaspectratio]{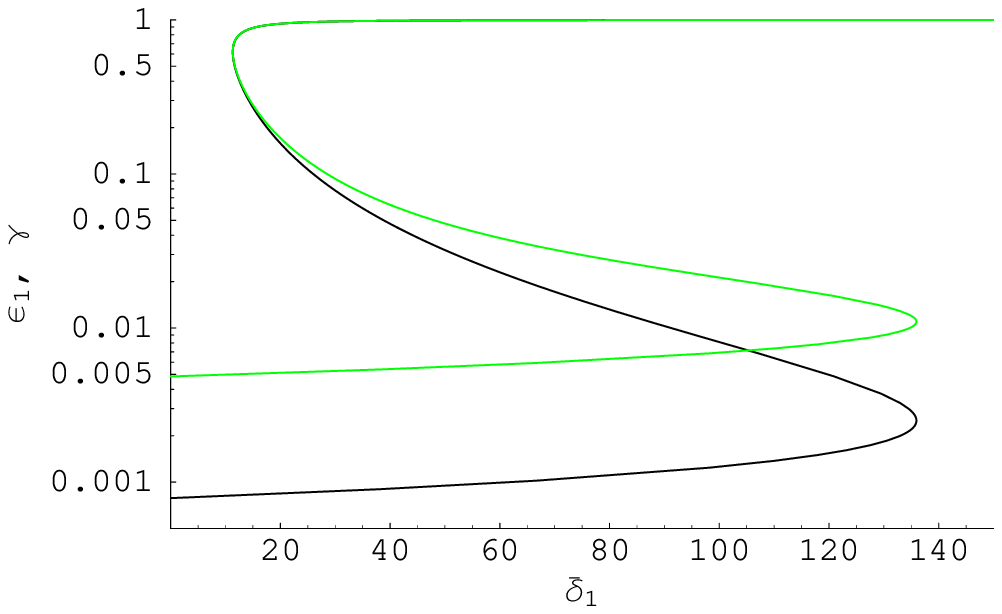}}\hspace*{0.1in}\subfigure[]{\includegraphics[width=3in,keepaspectratio]{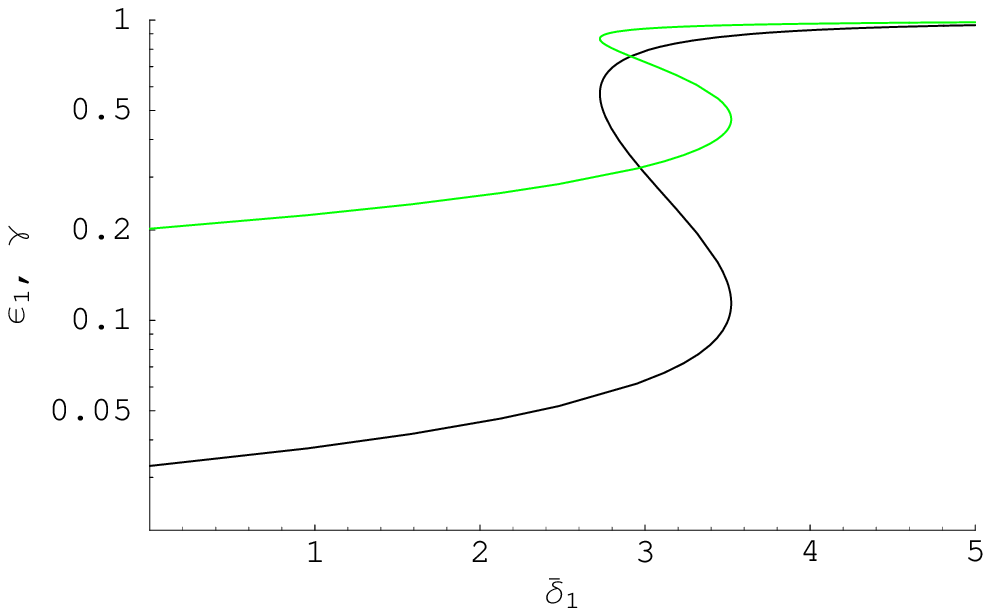}}\par\end{centering}

\caption{\label{fig:TMGbdWithLooping}\emph{Upper panel:} Bifurcation diagrams
for growth on TMG of (a)~wild-type cells ($\hat{\alpha}_{1}=1250$),
and (b)~low-repression cells ($\hat{\alpha}_{1}=30$). \emph{Lower
panel:} Variation of the steady state enzyme activity (black curve)
and fluorescence intensity (green curve) with the extracellular TMG
concentration ($\bar{\delta}_{1}$) for (c)~wild-type cells ($\hat{\alpha}_{1}=1250$,
$\alpha_{1}=20$), and (d)~low-repression cells ($\hat{\alpha}_{1}=30$,
$\alpha_{1}=0$).}
\end{figure}

In the presence of DNA looping, the steady states are given by the
equation\[
\frac{d\epsilon_{1}}{d\tau}=f(\epsilon)\equiv\frac{1}{1+\alpha_{1}/\left(1+\bar{\delta}_{1}\epsilon_{1}\right)^{2}+\hat{\alpha}_{1}/\left(1+\bar{\delta}_{1}\epsilon_{1}\right)^{4}}-\epsilon_{1}=0.\]
For each fixed $\hat{\alpha}_{1}\ge0$, this equation defines the
steady state surface in the $\alpha_{1}\bar{\delta}_{1}\epsilon_{1}$-space.
The folds on the steady state surface also satisfy the equation\[
f_{\epsilon_{1}}(\epsilon_{1})=\frac{2\alpha_{1}\bar{\delta}_{1}/\left(1+\bar{\delta}_{1}\epsilon_{1}\right)^{3}+4\hat{\alpha}_{1}\bar{\delta}_{1}/\left(1+\bar{\delta}_{1}\epsilon_{1}\right)^{5}}{\left[1+\alpha_{1}/\left(1+\bar{\delta}_{1}\epsilon_{1}\right)^{2}+\hat{\alpha}_{1}/\left(1+\bar{\delta}_{1}\epsilon_{1}\right)^{4}\right]^{2}}-1=0.\]
It is shown in Appendix~\ref{a:TMGbd} that the folds have the parametric
representation\begin{align*}
\alpha_{1}(\chi_{1}) & =\frac{1}{\left(1+\chi_{1}\right)^{2}}\frac{\hat{\alpha}_{1}\left(3\chi_{1}-1\right)-\left(1+\chi_{1}\right)^{5}}{\left(1-\chi_{1}\right)},\\
\epsilon_{1}(\chi_{1}) & =\frac{1}{2\chi_{1}}\frac{1-\chi_{1}}{\hat{\alpha}_{1}/\left(1+\chi_{1}\right)^{4}-1},\\
\bar{\delta}_{1}(\chi_{1}) & =\frac{\chi_{1}}{\epsilon_{1}(\chi_{1})}=2\chi_{1}^{2}\frac{\hat{\alpha}_{1}/\left(1+\chi_{1}\right)^{4}-1}{1-\chi_{1}}.\end{align*}
Furthermore, bistability is feasible at every $\hat{\alpha}_{1}\ge0$,
but there are three types of bifurcation diagrams:

\begin{enumerate}
\item If the repression due to DNA looping is small ($\hat{\alpha}_{1}<\left(5/3\right)^{5}\approx12.8$),
the bifurcation diagram is similar to the one obtained in the absence
of DNA looping: Bistability is feasible only if $\alpha_{1}$ is sufficiently
large; moreover, even if this condition is satisfied, bistability
occurs within a finite interval of $\bar{\delta}_{1}$ (Fig.~\ref{f:appTMG}a).
\item At intermediate levels of repression ($\left(5/3\right)^{5}<\hat{\alpha}_{1}<16$),
the bifurcation diagram contains two distinct bistable regions (Fig.~\ref{f:appTMG}b).
However, one of these regions is so small that it is unlikely to be
observed in practice.
\item If the repression due to DNA looping is large ($\hat{\alpha}_{1}>16$),
the bifurcation diagram is qualitatively different: Bistability is
feasible at every $\alpha_{1}\ge0$ (Figs.~\ref{fig:TMGbdWithLooping}a,b).
This reflects the fact that if the repression due to DNA looping is
sufficiently large, the induction rate is cooperative even if there
is no repression due to repressor-operator binding.
\end{enumerate}
Thus, for all practical purposes, there are only two types of types
of bifurcation diagrams.

The simulations are consistent with the data shown in Fig.~\ref{f:Oudenaarden}a.
In wild-type cells, bistability occurs over the range $10\lesssim\bar{\delta}_{1}\lesssim130$
(Fig.~\ref{fig:TMGbdWithLooping}c), which is in reasonable agreement
with the 10-fold range observed in the experiments (3--30~$\mu$M
in Fig.~\ref{f:Oudenaarden}a).

At parameter values corresponding to the cells transfected with the
\emph{lac} operator ($\alpha_{1}\approx0$, $\hat{\alpha}_{1}\approx30$),
there is bistability, but the range of extracellular concentrations
supporting bistability is so small that the system is practically
monostable (Fig.~\ref{fig:TMGbdWithLooping}d). However, the model
cannot be compared to the data shown in Fig.~\ref{f:Oudenaarden}b,
since it does not account for the diffusive flux of TMG, which is
significant at extracellular TMG concentrations exceeding 50~$\mu$M.

We can estimate $\bar{\delta}_{1,m}$ by comparing the model prediction
with the experimental data. Figs.~\ref{fig:TMGbdWithLooping}c and
\ref{f:Oudenaarden}a show that at the upper fold point, $\bar{\delta}_{1}\approx140$,
and $s_{1,0}\approx30$~$\mu$M, respectively. Since $K_{s,1}=680$~$\mu$M~\citep{ozbudak04}
\[
130\approx\bar{\delta}_{1}=\bar{\delta}_{1,m}\frac{30}{680+30}\Rightarrow\bar{\delta}_{1,m}\sim3100.\]
This estimate is consistent with the experimental data. Indeed, the
model implies at saturating concentrations of TMG, the intracellular
concentration of TMG is $K_{x}^{-1}\bar{\delta}_{1,m}$. Since $K_{x}^{-1}\sim7$
~$\mu$M~\citep[Fig.~4B]{Oehler2006}, the intracellular TMG concentration
at saturating conditions is 20~mM, which is in reasonable agreement
with the experimentally measured value of 15~mM \citep[Fig.~5]{Kepes1960}.

\subsection{Growth on lactose}

\subsubsection{No DNA looping}

In the presence of lactose, the steady states satisfy the equation
\begin{equation}
g(\epsilon_{1})\equiv\frac{1}{1+\alpha_{1}/\left(1+\delta_{1}\epsilon_{1}\right)^{2}}-\epsilon_{1}^{2}=0,\label{eq:Case2Ess}\end{equation}
which defines the steady state surface in the $\alpha_{1}\delta_{1}\epsilon_{1}$-space.

It is clear that no matter what the parameter values, there is at
least one steady state. Indeed, (\ref{eq:Case2Ess}) implies that
the net rate of enzyme synthesis, $g(\epsilon_{1})$, is positive
if $\epsilon_{1}=0$, and negative if $\epsilon_{1}$ is sufficiently
large. It follows that there is at least one $1/(1+\alpha_{1})<\epsilon_{1}<1$
at which $g(\epsilon_{1})$ is zero.

It turns out, however, there is exactly one steady state because multiple
steady states are impossible. To see this, observe that multiple steady
states are feasible only if the steady state surface contains singular
points, i.e., there are points $\alpha_{1},\delta_{1},\epsilon_{1}>0$
satisfying (\ref{eq:Case2Ess}) and \emph{}the necessary condition\begin{equation}
g_{\epsilon_{1}}(\epsilon_{1})=\frac{1}{\left[1+\alpha_{1}/\left(1+\delta_{1}\epsilon_{1}\right)^{2}\right]^{2}}\frac{2\alpha_{1}\delta_{1}}{\left(1+\delta_{1}\epsilon_{1}\right)^{3}}-2\epsilon_{1}=0.\label{eq:Case2Ebifn}\end{equation}
But there are no such points because $g_{\epsilon_{1}}<0$ at every
point on the steady state surface. Indeed, (\ref{eq:Case2Ess}) implies
that every point on the steady state surface satisfies the relations\[
\frac{1}{\left[1+\alpha_{1}\left(1+\delta_{1}\epsilon_{1}\right)^{2}\right]^{2}}=\epsilon_{1}^{4},\;\frac{\alpha_{1}}{\left(1+\delta_{1}\epsilon_{1}\right)^{2}}=\frac{1}{\epsilon_{1}^{2}}-1.\]
Substituting these relations in (\ref{eq:Case2Ebifn}) yields \begin{equation}
g_{\epsilon_{1}}(\epsilon_{1})=2\epsilon_{1}\left[\left(1-\epsilon_{1}^{2}\right)\frac{\delta_{1}\epsilon_{1}}{1+\delta_{1}\epsilon_{1}}-1\right]<0.\label{eq:NoBistabilityLactose}\end{equation}
It follows that there are no singular points on the steady state manifold,
and hence, no multiple steady states.

\begin{figure}[t]
\noindent \begin{centering}\includegraphics[width=3in,keepaspectratio]{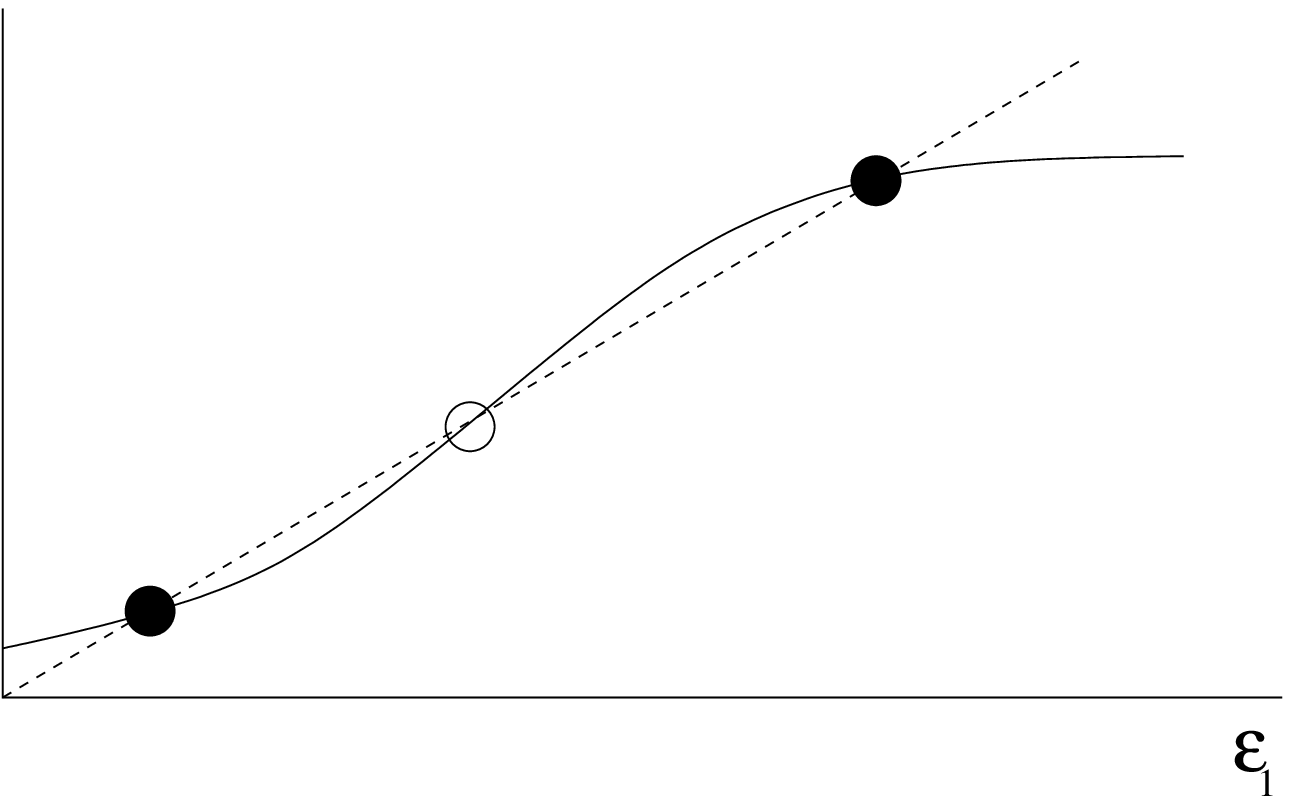}\hspace*{0.3in}\includegraphics[width=3in,keepaspectratio]{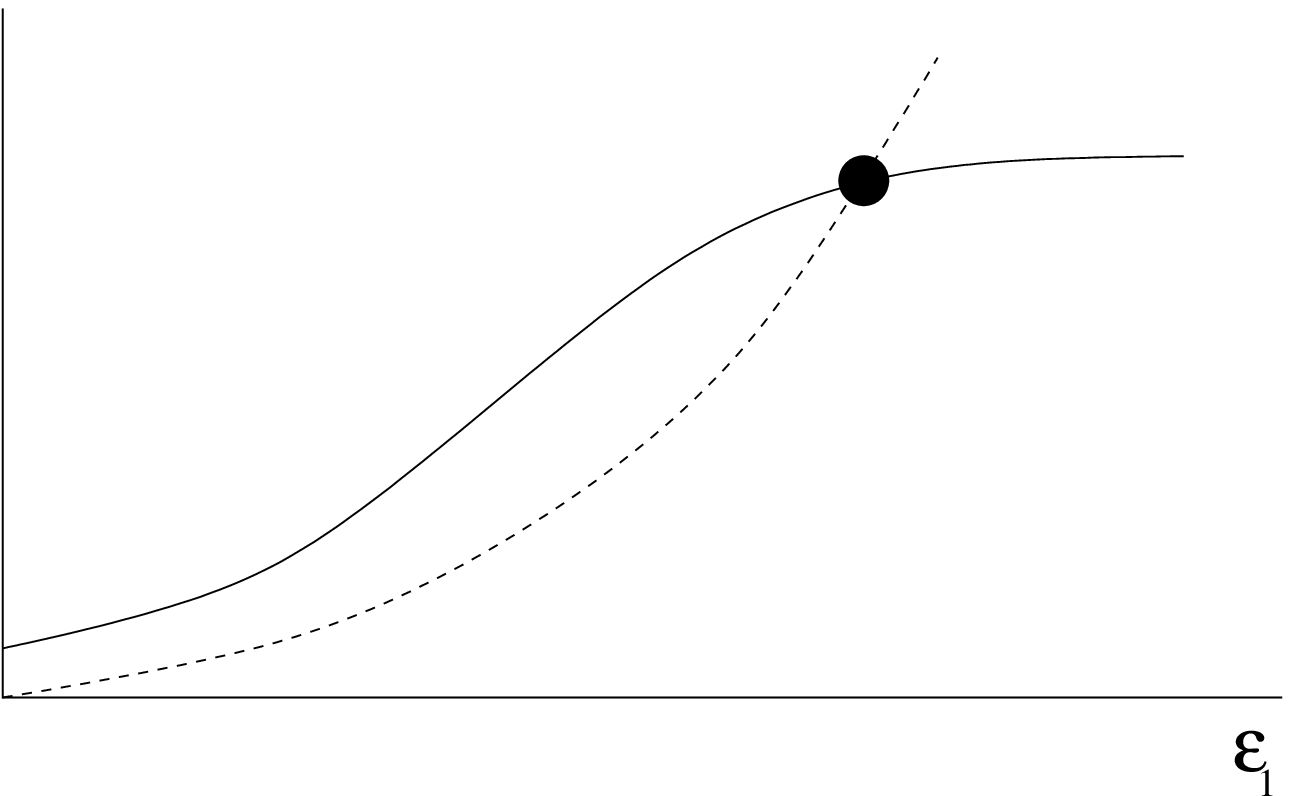}\par\end{centering}

\caption{\label{fig:LactoseIntuitive}Intuitive explanation for the absence
of bistability during growth on lactose of cells lacking DNA looping.
The enzyme synthesis and dilution rates are represented by full and
dashed lines, respectively. Stable and unstable steady states are
represented by full and open circles, respectively. (a)~During growth
on TMG, bistability is feasible because the dilution rate is proportional
to the enzyme level. (b)~During growth on lactose, bistability is
infeasible because the stabilizing effect of dilution ($\epsilon_{1}^{2}$)
is so strong that an unstable steady state (and hence, bistability)
is infeasible.}
\end{figure}

We pause for a moment to give the physical meaning of the foregoing
mathematical argument. To this end, observe that bistability can occur
only if the system permits the existence of an unstable steady state,
i.e., a steady state at which the slope of the induction rate exceeds
the slope of the dilution rate (Fig.~\ref{fig:LactoseIntuitive}a).%
\footnote{The stability of a steady state is completely determined by the relative
slopes of the induction and dilution rates at the steady state. If
the slope of the induction rate at a steady state exceeds the corresponding
slope of the dilution rate (see open circle in Fig.~\ref{fig:LactoseIntuitive}a),
the slightest increase (resp., decrease) in the enzyme level increases
(resp., decreases) the induction rate more than the dilution rate,
so that the enzyme level grows (resp., declines) even further.%
} Such a steady state is feasible when the cells are grown in the presence
of TMG because the dilution rate increases linearly with the enzyme
level. It is infeasible during growth on lactose because the dilution
rate, $\epsilon_{1}^{2}$, increases so rapidly with the enzyme level
that at every conceivable steady state, the slope of the dilution
rate exceeds the slope of the induction rate, i.e., $g_{\epsilon_{1}}<0$
at every possible steady state (Fig.~\ref{fig:LactoseIntuitive}b).
In other words, the stabilizing effect of dilution is so strong that
an unstable stable steady state, and hence, bistability, is impossible.

The model assumes that the yield is constant. It is well known, however,
that the yield is vanishingly small under starvation conditions, and
increases progressively as the nutritional status of the cells improves~\citep{tempest67}.
Thus, it seems plausible to assume that at low enzyme levels, the
lactose transport rate and yield are relatively small; as the enzyme
level increases, so do the lactose transport rate and the yield. If
this is true, it is relevant to ask if multistability is impossible
even if the yield is an increasing function of the enzyme level, i.e.,
the yield has the form $Y_{1}\phi(\epsilon_{1})$, where $\phi(\epsilon_{1})\le1$
is some increasing function of $\epsilon_{1}$. The foregoing physical
explanation suggests that multistability is, \emph{a fortiori}, impossible
because the dilution rate now increases with $\epsilon_{1}$ at a
rate even faster than $\epsilon_{1}^{2}$. Analysis confirms this
intuitive argument --- the value of $g_{\epsilon_{1}}$ at any point
on the steady state surface is \[
2\phi(\epsilon_{1})\epsilon_{1}\left[\left\{ 1-\phi(\epsilon_{1})\epsilon_{1}^{2}\right\} \frac{\delta_{1}\epsilon_{1}}{1+\delta_{1}\epsilon_{1}}-1\right]-\phi_{\epsilon_{1}}(\epsilon_{1})\epsilon_{1}^{2}<0.\]
Thus, multistability is impossible even if the yield is an increasing
function of the enzyme level.

The above argument shows that the existence of bistability is not
determined solely by the intensity of the destabilizing positive feedback
generated by induction. It also depends on the strength the stabilizing
effect of dilution. If this stabilizing effect is sufficiently large,
it can neutralize the destabilizing effect of positive feedback.

\subsubsection{DNA looping}

We have shown above that in the absence of DNA looping, there is no
bistability during growth on lactose because the dilution rate, which
is proportional to $\epsilon_{1}^{2}$, neutralizes the destabilizing
effect of positive feedback. However, in the presence of DNA looping,
the destabilizing effect of positive feedback is much stronger because
the induction rate contains terms proportional to $\epsilon_{1}^{4}$.
Under these conditions, it seems plausible to expect that bistability
is feasible at sufficiently large $\hat{\alpha}_{1}$. We show below
that is indeed the case, but the bistable region is extremely small.

\begin{figure}[t]
\noindent \begin{centering}\subfigure[]{\includegraphics[width=3in,keepaspectratio]{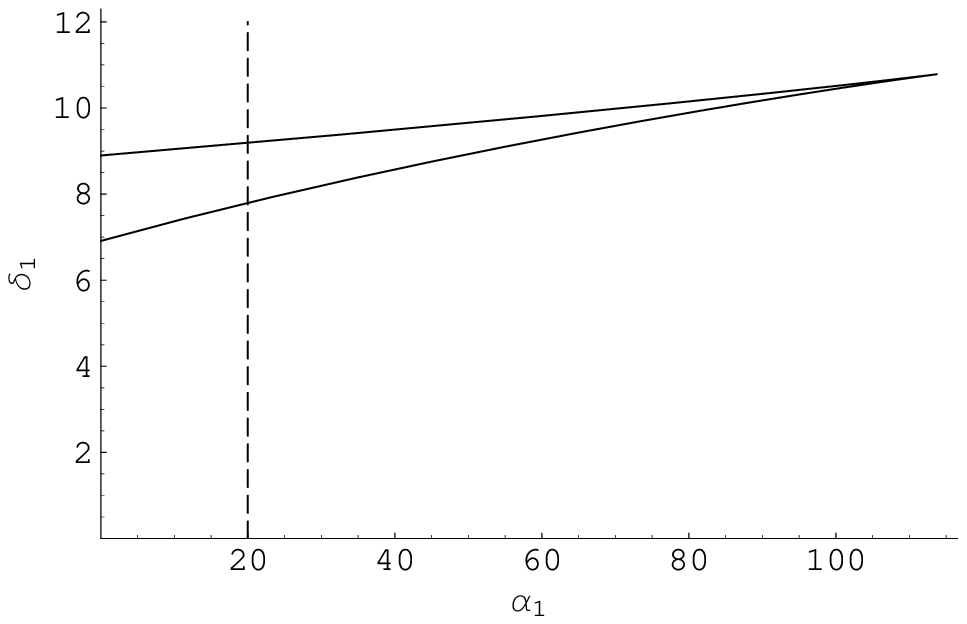}}\hspace*{0.3in}\subfigure[]{\includegraphics[width=3in,keepaspectratio]{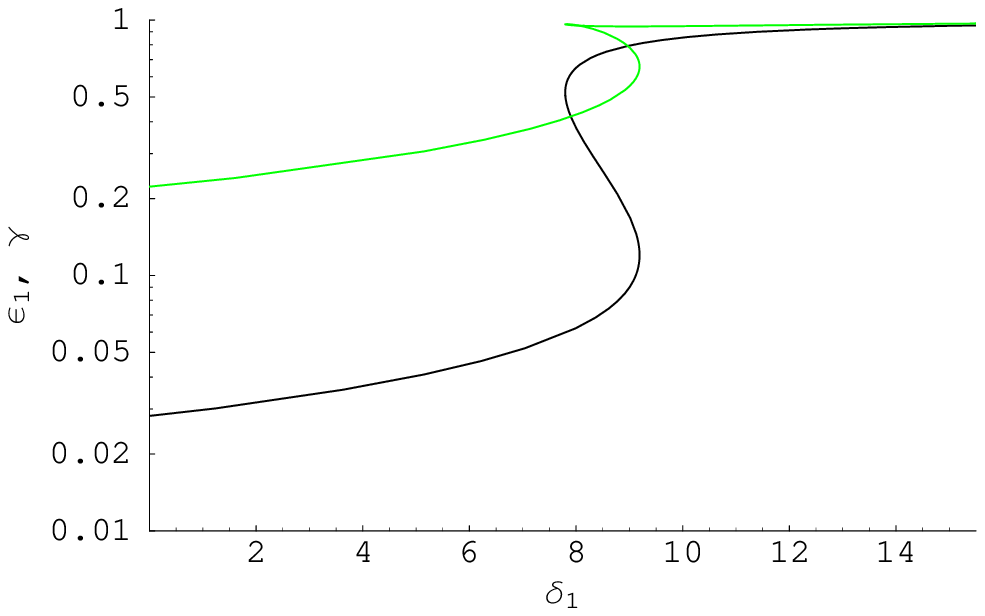}}\par\end{centering}

\caption{\label{fig:bdLactose}Growth of wild-type cells on lactose: (a) Bifurcation
diagram ($\hat{\alpha}_{1}=1250$). The dashed line shows the path
along which $\delta_{1}$ changes when the extracellular lactose concentration
is varied. (b) The variation of the steady state enzyme activity (black
curve) and fluorescence intensity (green curve) with $\delta_{1}$,
a surrogate for the extracellular lactose concentration ($\hat{\alpha}_{1}=1250$,
$\alpha_{1}=20$).}
\end{figure}

In the presence of DNA looping, the bifurcation points satisfy the
equations\begin{align*}
g(\epsilon_{1}) & \equiv\frac{1}{1+\alpha_{1}/\left(1+\delta_{1}\epsilon_{1}\right)^{2}+\hat{\alpha}_{1}/\left(1+\delta_{1}\epsilon_{1}\right)^{4}}-\epsilon_{1}^{2}=0,\\
g_{\epsilon_{1}}(\epsilon_{1}) & =\frac{2\alpha_{1}\delta_{1}/\left(1+\delta_{1}\epsilon_{1}\right)^{3}+4\hat{\alpha}_{1}\delta_{1}/\left(1+\delta_{1}\epsilon_{1}\right)^{5}}{\left[1+\alpha_{1}/\left(1+\delta_{1}\epsilon_{1}\right)^{2}+\hat{\alpha}_{1}/\left(1+\delta_{1}\epsilon_{1}\right)^{4}\right]^{2}}-2\epsilon_{1}=0,\end{align*}
For each fixed $\hat{\alpha}_{1}>0$, these two equations define the
locus of the bifurcation points in $\alpha_{1}\delta_{1}\epsilon_{1}$-space.
It is shown in Appendix~\ref{a:LactoseBD} that this curve has the
parametric representation\begin{align*}
\epsilon_{1}(\chi_{1}) & =\sqrt{\frac{1}{\chi_{1}\left\{ \hat{\alpha}_{1}/\left(1+\chi_{1}\right)^{4}-1\right\} }},\\
\alpha_{1}(\chi_{1}) & =\frac{\hat{\alpha}_{1}\left(\chi_{1}-1\right)-\left(1+\chi_{1}\right)^{5}}{\left(1+\chi_{1}\right)^{2}},\\
\delta_{1}(\chi_{1}) & =\frac{\chi_{1}}{\epsilon_{1}(\chi_{1})}=\chi_{1}^{3/2}\sqrt{\frac{\hat{\alpha}_{1}}{\left(1+\chi_{1}\right)^{4}}-1}.\end{align*}
Furthermore, bistability is feasible only if $\hat{\alpha}_{1}>5^{5}/2^{4}\approx195$.

In wild-type cells, for instance, bistability occurs for all $7\lesssim\delta_{1}\lesssim9$
(Fig.~\ref{fig:bdLactose}a). Now, it is conceivable that bistability
is not observed during growth on lactose because the strength of the
positive feedback, $\delta_{1,m}$, is so small (<7) that $\delta_{1}\equiv\delta_{1,m}\sigma_{1}$remains
below the bistable region at all concentrations of extracellular lactose.
It is clear, however, that even if $\delta_{1,m}$ is large (>9),
the bistable region is so narrow that it is unlikely to be observed
in practice (Fig.~\ref{fig:bdLactose}b). We show below that the
width of the bistable region increases dramatically in the presence
of glucose.

In wild-type cells, bistability is feasible during growth on lactose,
but the bifurcation diagram is qualitatively different from that obtained
during growth on TMG. Indeed, during growth on TMG, bistability is
feasible for all $\alpha_{1}\ge0$ (Fig.~\ref{fig:TMGbdWithLooping}a).
In sharp contrast, bistability is feasible during growth on lactose
only if $\alpha_{1}$ is sufficiently small (Fig.~\ref{fig:bdLactose}a).
The qualitatively different bifurcation diagram for lactose can be
explained as follows. During growth on lactose, bistability can occur
only if DNA looping has a strong effect on the induction rate. But
DNA looping dominates the induction rate precisely if \[
\frac{\hat{\alpha}_{1}}{\left(1+\chi_{1}\right)^{4}}\gg\frac{\alpha_{1}}{\left(1+\chi_{1}\right)^{2}}\Rightarrow\chi_{1}\ll\sqrt{\frac{\hat{\alpha}_{1}}{\alpha_{1}}}-1.\]
It follows that regardless of the value of $\hat{\alpha}_{1}$, the
range of inducer concentrations at which DNA looping is dominant vanishes
at a sufficiently high value of $\alpha_{1}$. Consequently, the induction
rate, and hence, the dynamics, become similar to those observed in
the absence of DNA looping.

\subsection{Growth on lactose + glucose}

In the presence of glucose and lactose, the steady states satisfy
the equations\begin{align}
g_{1}(\epsilon_{1},\epsilon_{2}) & \equiv\frac{1}{1+\alpha_{1}/\left(1+\delta_{1}\epsilon_{1}\right)^{2}+\hat{\alpha}_{1}/\left(1+\delta_{1}\epsilon_{1}\right)^{4}}-\left(\epsilon_{1}+\alpha\epsilon_{2}\right)\epsilon_{1}=0,\label{eq:GLe1}\\
g_{2}(\epsilon_{1},\epsilon_{2}) & \equiv\frac{\alpha}{1+\alpha_{2}/\left(1+\delta_{2}\epsilon_{2}\right)}-\left(\epsilon_{1}+\alpha\epsilon_{2}\right)\epsilon_{2}=0,\label{eq:GLe2}\end{align}
Our goal is to address the following question: Given a particular
cell type, what are concentrations of lactose and glucose at which
bistability is feasible? This question is difficult to address because
there are three parameters that depend on the substrate concentrations
($\delta_{i}\propto\sqrt{\sigma_{i}}$, $\alpha\propto\sqrt{\sigma_{2}/\sigma_{1}}$).
Since the combination of parameters \[
\beta\equiv\frac{\alpha\delta_{1}}{\delta_{2}}=\frac{Y_{2}k_{x,2}^{+}K_{x,2}^{-1}}{Y_{1}k_{x,1}^{+}K_{x,1}^{-1}}\]
is independent of the substrate concentrations, it is convenient to
replace $\alpha$ with $\beta\delta_{2}/\delta_{1}$. In terms of
the model, the question of interest then becomes: Given any fixed
$\alpha_{1},\hat{\alpha}_{2},\alpha_{2},\beta$, what are the values
of $\delta_{1}$ and $\delta_{2}$ at which bistability is feasible?
Unlike $\alpha_{1},\hat{\alpha}_{2},\alpha_{2}$, the value of $\beta$
cannot be determined from the experimental literature, since $K_{x,2}^{-1}$
is a phenomenological parameter. In the simulations, we assume that
$\beta=1$. However, we show below that this is not a particularly
restrictive assumption, since the qualitative behavior of the bifurcation
diagram is completely determined by $\alpha_{1}$ and $\hat{\alpha}_{1}$.

\begin{figure}
\noindent \begin{centering}\includegraphics[width=3in,keepaspectratio]{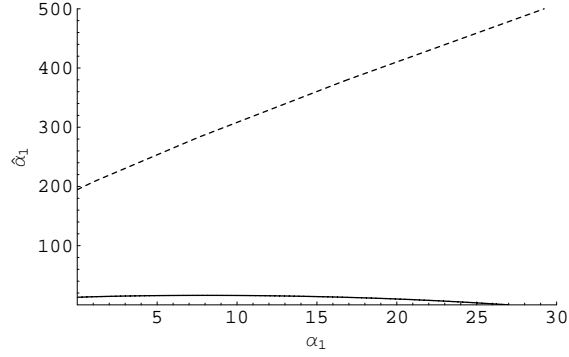}\par\end{centering}

\caption{\label{f:bdA1A2H}Classification of the dynamics during growth on
lactose + glucose. Bistability is feasible if and only if $\alpha_{1}$
and $\hat{\alpha}_{1}$ lie above the full curve. If $\alpha_{1}$
and $\hat{\alpha}_{1}$ lie between the full and dashed curves (resp.,
aboe the dashed curve), the bifurcation diagram has the form shown
in Fig.~\ref{f:GLbdPlotNoLooping} (resp., Fig.~\ref{f:GLbdPlotLooping}a).}
\end{figure}

\begin{figure}
\noindent \begin{centering}\subfigure[]{\includegraphics[width=3in,keepaspectratio]{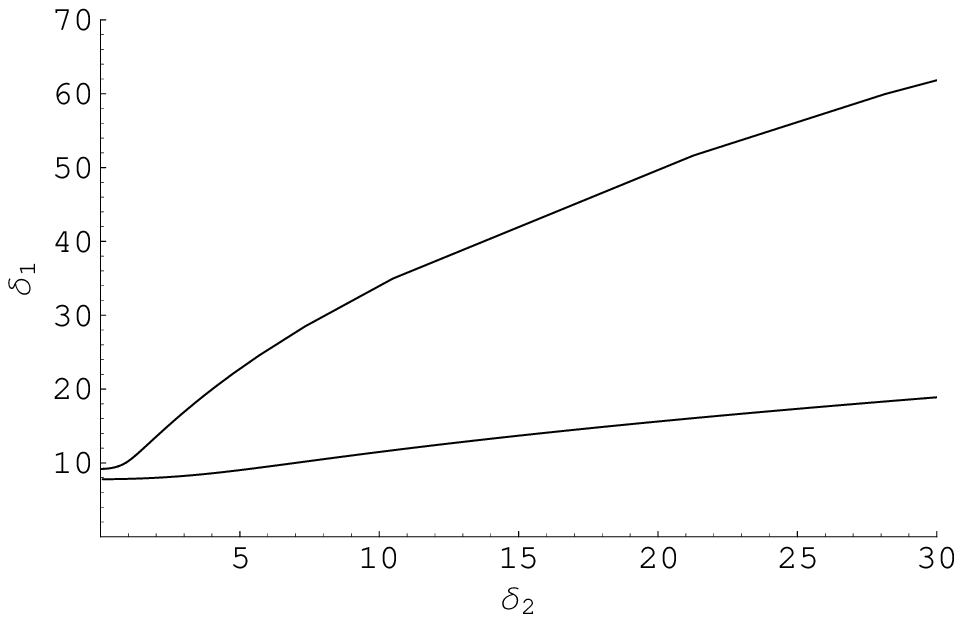}}\hspace*{0.3in}\subfigure[]{\includegraphics[width=3in,keepaspectratio]{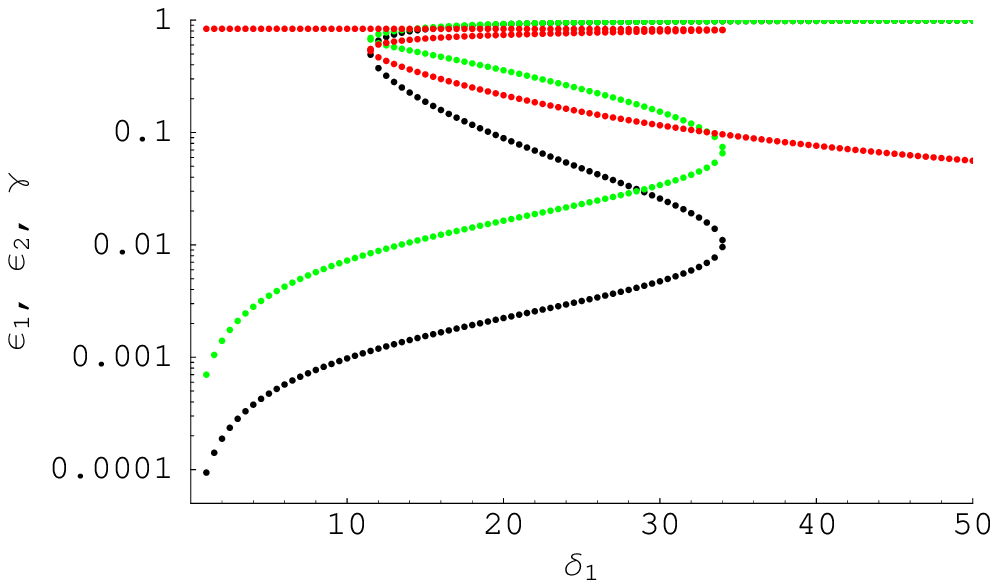}}\par\end{centering}

\caption{\label{f:GLbdPlotLooping}Steady states of wild-type cells during
growth on glucose and lactose ($\alpha_{1}=20$, $\hat{\alpha}_{1}=1250$,
$\alpha_{2}=4$, $\beta=1$). (a)~Bifurcation diagram. Bistability
occurs if $\delta_{1},\delta_{2}$ lie in the region between the curves.
(b)~Variation of the steady state activities of $E_{1}$ (black curve),
$E_{2}$ (red curve) and $G$ (green curve) with the extracellular
lactose concentration, $\delta_{1}$, at the fixed glucose concentration,
$\delta_{2}=10$.}
\end{figure}

Bistability is feasible only if there are steady states satisfying
(\ref{eq:GLe1})--(\ref{eq:GLe2}) and the condition\[
\det\left[\begin{array}{cc}
\frac{\partial g_{1}}{\partial\epsilon_{1}} & \frac{\partial g_{1}}{\partial\epsilon_{2}}\\
\frac{\partial g_{2}}{\partial\epsilon_{1}} & \frac{\partial g_{2}}{\partial\epsilon_{2}}\end{array}\right]=0.\]
It is shown in Appendix~\ref{a:GLbd} that these three equations
can be rewritten as \begin{align}
\delta_{1}(\chi_{1},\chi_{2}) & =\chi_{1}\sqrt{1+\alpha_{1}/\left(1+\chi_{1}\right)^{2}+\hat{\alpha}_{1}/\left(1+\chi_{1}\right)^{4}}\sqrt{\frac{1}{1+h(\chi_{1},\chi_{2})}},\label{eq:GLdelta1}\\
\delta_{2}(\chi_{1},\chi_{2}) & =\chi_{2}\sqrt{1+\alpha_{2}/\left(1+\chi_{2}\right)}\sqrt{\frac{1+h(\chi_{1},\chi_{2})}{h(\chi_{1},\chi_{2})}},\label{eq:GLdelta2}\\
\beta & =\frac{\chi_{1}}{\chi_{2}}h(\chi_{1},\chi_{2}),\label{eq:GLbeta}\end{align}
where\begin{align*}
h(\chi_{1},\chi_{2}) & \equiv\frac{1/p(\chi_{1})-1}{1-q(\chi_{2})},\\
p(\chi_{1}) & =2\frac{\alpha_{1}/\left(1+\chi_{1}\right)^{2}+2\hat{\alpha}_{1}/\left(1+\chi_{1}\right)^{4}}{1+\alpha_{1}/\left(1+\chi_{1}\right)^{2}+\hat{\alpha}_{1}/\left(1+\chi_{1}\right)^{4}}\frac{\chi_{1}}{1+\chi_{1}}-1,\\
q(\chi_{2}) & =\frac{\alpha_{2}/\left(1+\chi_{2}\right)}{1+\alpha_{2}/\left(1+\chi_{2}\right)}\frac{\chi_{2}}{1+\chi_{2}}-1.\end{align*}
Furthermore, bistability is feasible if and only if $\alpha_{1}$
and $\hat{\alpha}_{1}$ lie above the full curve shown in Fig.~\ref{f:bdA1A2H}.
It is precisely for such values of $\alpha_{1}$ and $\hat{\alpha}_{1}$
that eq.~(\ref{eq:GLbeta}) has positive solutions, which determine
a curve in the $\chi_{1}\chi_{2}$-space. The variation of $\delta_{1}(\chi_{1},\chi_{2})$
and $\delta_{2}(\chi_{1},\chi_{2})$ along this curve circumscribes
the bistability region on the $\delta_{1}\delta_{2}$-plane (for the
cell type defined by the fixed values of $\alpha_{1},\hat{\alpha}_{1},\alpha_{2},\beta$).
If $\alpha_{1}$ and $\hat{\alpha}_{1}$ lie between the full and
dashed curves of Fig.~\ref{f:bdA1A2H}, the bifurcation curve has
the form shown in Fig.~\ref{f:GLbdPlotNoLooping} --- it does not
intersect the $\delta_{1}$-axis. If $\alpha_{1}$ and $\hat{\alpha}_{1}$
lie above the dashed curve of Fig.~\ref{f:bdA1A2H}, the bifurcation
curves intersect the $\delta_{1}$-axis (Fig.~\ref{f:GLbdPlotLooping}a).
Thus, the form of the bifurcation curve is completely determined by
$\alpha_{1}$ and $\hat{\alpha}_{1}$.

Fig.~\ref{f:bdA1A2H} implies that in cells lacking DNA looping,
bistability is feasible if and only if $\hat{\alpha}_{1}$ exceeds
the threshold value of 27, which is identical to the threshold in
such cells when they grow in the presence of TMG~(Fig.~\ref{f:TMGbdNoLooping}).
The existence of this identity is not coincidental. As we show below,
the dynamics of growth on lactose + glucose are, in some sense, identical
to the dynamics of growth in the presence of TMG.

Fig.~\ref{f:GLbdPlotLooping}a shows the bifurcation diagram for
wild-type cells growing on lactose + glucose. Evidently, bistability
is feasible at every $\delta_{2}\ge0$. This mirrors the fact that
during growth of wild-type cells on pure lactose ($\delta_{2}=0$),
bistability is feasible, although the range of lactose concentrations
supporting bistability is very small. At relatively modest values
of the extracellular glucose concentration, $\delta_{2}$, the bistable
region increases dramatically. This is because at sufficiently large
concentrations of glucose, the dilution rate due to growth on glucose
becomes significant. Importantly, this dilution rate is formally similar
to the dilution rate during growth on TMG, insofar as it increases
linearly with $\epsilon_{1}$. Thus, the dynamics are also analogous
to those observed during growth on TMG: Bistability is feasible for
a range of lactose concentrations ($\delta_{1}$).

Fig.~\ref{f:GLbdPlotLooping}a is qualitatively similar to the bifurcation
diagram obtained by Santillan et al~\citep[Fig.~2]{Santillan2007}.
However, in their model, the width of the bistability region increases
in the presence of glucose due to regulatory effects, such as cAMP
activation and inducer exclusion, exerted on the induction rate of
the \emph{lac} operon. In our model, the same phenomenon occurs because
the stabilizing effect of dilution is significantly smaller during
growth on lactose + glucose (as opposed to growth on pure lactose).

It remains to explain the discrepancy between the data obtained Ozbudak
et al and Loomis \& Magasanik. The simplest explanation is that in
the strain used by Ozbudak et al, the parameter, $\delta_{1,m}$,
which measures the strength of the positive feedback, is so small
(<7) that $\delta_{1}\equiv\delta_{1,m}\sigma_{1}$ does not enter
the bistability region at any extracellular lactose concentration
(Fig.~\ref{f:GLbdPlotLooping}a). On the other hand, the values of
$\delta_{1,m},\delta_{2,m}$ for the strain used by Loomis \& Magasanik
are so large that $\delta_{1},\delta_{2}$ lie in the bistable region
at the extracellular lactose and glucose concentrations used in their
experiments.

It should be noted, however, that the discrepancy could also reflect
technical differences in the experiments. Fig~\ref{f:GLbdPlotLooping}b
shows the steady state levels of the lactose enzymes (black curve),
glucose enzymes (red curve), and GFP (green curve). In the neighborhood
of the upper bifurcation point ($\delta_{1}\approx35$), the lactose
enzyme level of the induced cells is $\sim$100 times the lactose
enzyme levels of the non-induced cells. However, the GFP level of
the induced cells is no more than $\sim$ 10--20 fold higher than
the GFP level of the non-induced cells. Now, Fig.~\ref{f:Oudenaarden}a
shows that in the neighborhood of the upper bifurcation point, the
steady state fluorescence of the non-induced cells is scattered over
a $\sim$50-fold range (see the fluorescence distribution in lower
panel of Fig.~\ref{f:Oudenaarden}a over the range 20--30~$\mu$M
). Assuming that the distribution has a similar variance during growth
on lactose, it is conceivable that even if bistability exists, the
fluorescence distributions of the non-induced and induced cells overlap,
and appear to be unimodal.

In Fig~\ref{f:GLbdPlotLooping}b, the precise values of the \emph{lac}
enzyme and GFP levels in induced and non-induced cells depend on the
particular choice of the parameter values. However, as shown above,
the ratio of GFP levels in induced and non-induced cells is always
smaller than the corresponding ratio for the \emph{lac} enzymes. Experimental
artefacts indicating the absence of bistability are therefore more
likely if GFP levels, rather than enzyme activities, are measured.

\section{Conclusions}

The experimental data shows that the existence of bistability in the
\emph{lac} operon depends on the composition of nutrient medium. It
occurs during growth on TMG/succinate, but not on lactose. There are
conflicting reports of its existence in media containing lactose and
glucose.

The occurrence or absence of bistability reflects the net result of
the destabilizing effect due to the positive feedback generated by
induction and the stabilizing effect of dilution. In previous models,
the experimental data has been rationalized entirely in terms of changes
in the strength of positive feedback as a function of the medium composition.
We have shown above that:

\begin{enumerate}
\item The stabilizing effect of dilution also changes dramatically with
the composition of the medium. In the presence of non-galactosidic
carbon sources, such as succinate or glucose, the dilution rate of
the \emph{lac} enzymes contains a term that is proportional to the
activity of these enzymes. During growth on pure lactose, the dilution
rate of \emph{lac} enzymes is proportional to the square of their
activity.
\item These variations in the functional form of the dilution rate have
a profound effect on the dynamics. During growth on lactose, the stabilizing
effect of dilution is so strong that bistability is virtually impossible
even if induction is subject to the strong positive feedback generated
by DNA looping. During growth on TMG/succinate or lactose + glucose,
bistability is feasible because the stabilizing effect of dilution
decreases sharply. Thus, bistabililty is much more likely in the presence
of non-galactosidic carbon sources, namely, succinate and glucose.
\item The conflicting results on bistability during growth on lactose +
glucose can be explained in terms of the relative magnitudes of the
destabilizing and stabilizing effects of positive feedback and dilution,
respectively. However, the criterion used by Ozbudak et al to infer
monostability, namely, the absence of a bimodal distribution in non-induced
cells, is prone to error. Since the repression of the reporter \emph{lac}
operon is much lower than the repression of the native \emph{lac}
operon, the ratio of GFP levels in induced and non-induced cells is
significantly lower than the corresponding ratio of the enzyme activities.
The fluorescence distribution can therefore appear to be unimodal
even if the enzyme levels are bistable.
\end{enumerate}
Taken together, these results show that while the intensity of the
positive feedback undoubtedly influences the dynamics of the \emph{lac}
operon, the dilution rate also has profound effects. These effects
can be discerned only if it is recognized that the specific growth
rate is not necessarily a fixed parameter --- it depends on the physiological
state of the cells.

\noindent \begin{flushleft}\textbf{Acknowledgment:}\par\end{flushleft}

This research was supported in part with funds from the National Science
Foundation under contract NSF DMS-0517954. We are grateful to the
anonymous reviewers for their valuable comments.

\appendix

\section{\label{a:GlucoseLactoseEqns}Derivation of eqs.~(\ref{eq:MSe1Orig1})--(\ref{eq:MSxOrig1})}

Since $x_{1}+x_{2}+e_{1}+g+e_{2}+c^{-}=1$, addition of eqs.~(\ref{eq:MSxOrig})--(\ref{eq:MScMinusOrig})
yields\[
0=r_{g}-\frac{1}{c}\frac{dc}{dt},\]
where\[
r_{g}\equiv\sum_{i=1}^{2}\left[V_{s,i}e_{i}\frac{s_{i}}{K_{s,i}+s_{i}}-k_{x,i}^{-}x_{i}-\left(1-Y_{i}\right)k_{x,i}^{+}x_{i}\right].\]
is the specific growth rate. This becomes evident if we rewrite the
above equation in the more familiar form\[
\frac{dc}{dt}=r_{g}c.\]
Alternatively, one can see that as expected, $r_{g}$ is the \emph{net}
rate of uptake of the two substrates (uptake minus loss by excretion
and respiration).

If the experiments are started with extremely small inocula and terminated
before the cell densities become significantly large, the substrate
concentrations do not change significantly over the course of the
experiment, i.e. $s_{i}(t)\approx s_{i,0}\equiv s_{i}(0)$ for all
$t$. Since $k_{x,i}^{+}>k_{x,i}^{-}$ is large, $x_{i}$ rapidly
attains quasisteady state, i.e., eq.~(\ref{eq:MSxOrig}) becomes\[
0\approx V_{s,i}e_{i}\frac{s_{i,0}}{K_{s,i}+s_{i,0}}-k_{x,i}^{-}x_{i}-k_{x,i}^{+}x_{i},\]
which implies that \begin{align}
x_{i} & \approx\frac{V_{s,i}}{k_{x,i}^{+}+k_{x,i}^{-}}e_{i}\sigma_{i},\;\sigma_{i}\equiv\frac{s_{i,0}}{K_{s,i}+s_{i,0}},\label{eq:MSredX}\\
r_{g} & \approx\sum_{i=1}^{2}Y_{i}k_{x,i}^{+}x_{i}=\sum_{j=1}^{2}Y_{i}\phi_{i}V_{s,i}e_{i}\sigma_{i},\;\phi_{i}\equiv\frac{k_{x,i}^{+}}{k_{x,i}^{-}+k_{x,i}^{+-}}.\label{eq:MSredrG}\end{align}
Eq.~(\ref{eq:MSredX}) is identical to eq.~(\ref{eq:MSxOrig1}).
Since $(1/c)(dc/dt)=r_{g}$, substituting (\ref{eq:MSredrG}) in (\ref{eq:MSe1Orig})--(\ref{eq:MSe2Orig})
yields (\ref{eq:MSe1Orig1})--(\ref{eq:MSe2Orig1}).

\section{\label{a:TMGbd}Bifurcation diagram for growth in the presence of
TMG}

The bifurcation points satisfy the equations\begin{align*}
f(\epsilon_{1}) & \equiv\frac{1}{1+\alpha_{1}/\left(1+\bar{\delta}_{1}\epsilon_{1}\right)^{2}+\hat{\alpha}_{1}/\left(1+\bar{\delta}_{1}\epsilon_{1}\right)^{4}}-\epsilon_{1}=0,\\
f_{\epsilon_{1}}(\epsilon_{1}) & =\frac{2\alpha_{1}\bar{\delta}_{1}/\left(1+\bar{\delta}_{1}\epsilon_{1}\right)^{3}+4\hat{\alpha}_{1}\bar{\delta}_{1}/\left(1+\bar{\delta}_{1}\epsilon_{1}\right)^{5}}{\left[1+\alpha_{1}/\left(1+\bar{\delta}_{1}\epsilon_{1}\right)^{2}+\hat{\alpha}_{1}/\left(1+\bar{\delta}_{1}\epsilon_{1}\right)^{4}\right]^{2}}-1=0,\end{align*}
which imply that\begin{align*}
\frac{\alpha_{1}}{\left(1+\bar{\delta}_{1}\epsilon_{1}\right)^{2}}+\frac{\hat{\alpha}_{1}}{\left(1+\bar{\delta}_{1}\epsilon_{1}\right)^{4}} & =\frac{1}{\epsilon_{1}}-1,\\
\frac{\alpha_{1}}{\left(1+\bar{\delta}_{1}\epsilon_{1}\right)^{2}}+2\frac{\hat{\alpha}_{1}}{\left(1+\bar{\delta}_{1}\epsilon_{1}\right)^{4}} & =\frac{1}{2}\left(1+\frac{1}{\bar{\delta}_{1}\epsilon_{1}}\right)\frac{1}{\epsilon_{1}}.\end{align*}
Since $\chi_{1}\equiv\bar{\delta}_{1}\epsilon_{1}$, we have two equations
that are linear with respect to $\alpha_{1}$, $\hat{\alpha}_{1}$,
and $1/\epsilon_{1}$. Solving for $\alpha_{1}$ and $1/\epsilon_{1}$
in terms of $\hat{\alpha}_{1}$ and $\chi_{1}$ yields\begin{align*}
\alpha_{1}(\chi_{1}) & =\frac{1}{\left(1+\chi_{1}\right)^{2}}\frac{\psi_{1}(\chi_{1})}{1-\chi_{1}},\;\psi_{1}(\chi_{1})\equiv\hat{\alpha}_{1}\left(3\chi_{1}-1\right)-\left(1+\chi_{1}\right)^{5},\\
\epsilon_{1}(\chi_{1}) & =\frac{1}{2\chi_{1}}\frac{1-\chi_{1}}{\psi_{2}(\chi_{1})},\;\psi_{2}(\chi_{1})\equiv\frac{\hat{\alpha}_{1}}{\left(1+\chi_{1}\right)^{4}}-1,\\
\bar{\delta}_{1}(\chi_{1}) & =\frac{\chi_{1}}{\epsilon_{1}(\chi_{1})}=2\chi_{1}^{2}\frac{\psi_{2}(\chi_{1})}{1-\chi_{1}}.\end{align*}
These three equations provide a parametric representation of the bifurcation
curve.

In the absence of DNA looping, the parametric representation of the
curve becomes\[
\epsilon_{1}(\chi_{1})=\frac{\chi_{1}-1}{2\chi_{1}},\;\alpha_{1}(\chi_{1})=\frac{\left(1+\chi_{1}\right)^{3}}{\chi_{1}-1},\;\bar{\delta}_{1}(\chi_{1})=\frac{2\chi_{1}^{2}}{\chi_{1}-1}.\]
It follows that the bifurcation curve exists (i.e. lies in the positive
octant of the $\alpha_{1}\bar{\delta}_{1}\epsilon_{1}$-space) for
all $\chi_{1}>1$.

\begin{figure}
\noindent \begin{centering}\subfigure[]{\includegraphics[width=3in,keepaspectratio]{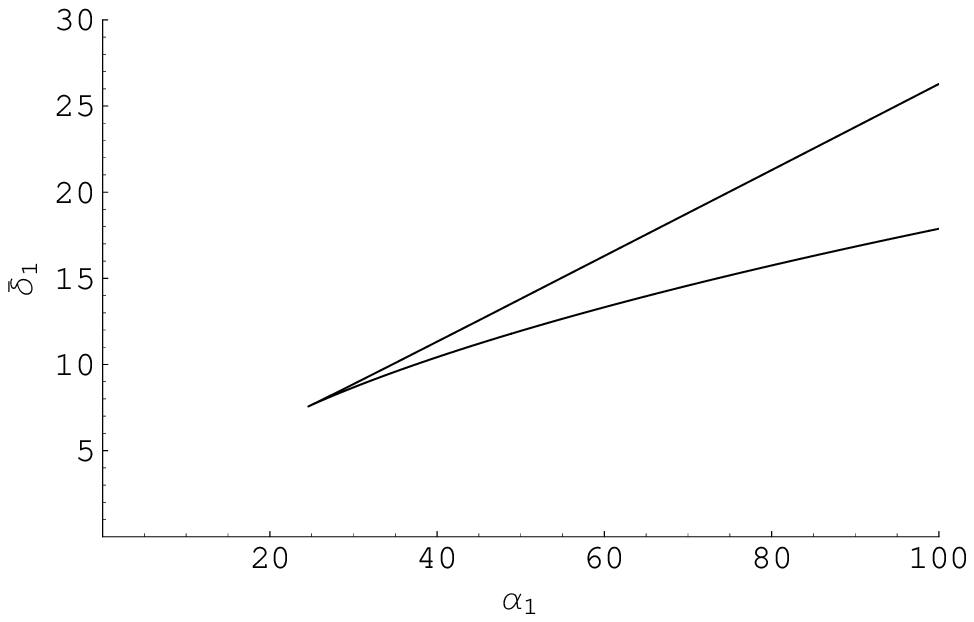}}\hspace*{0.2in}\subfigure[]{\includegraphics[width=3in,keepaspectratio]{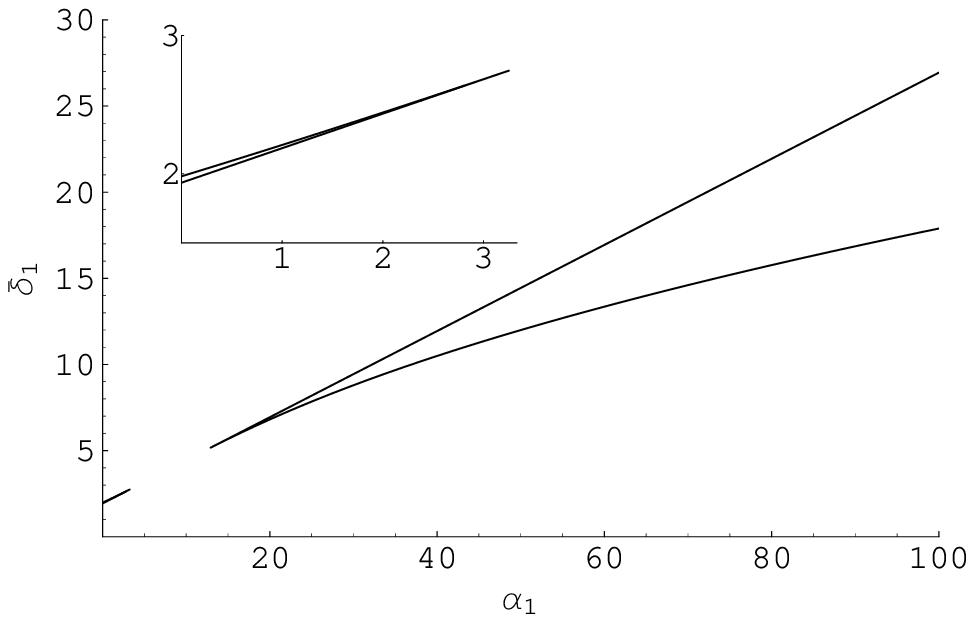}}\par\end{centering}

\caption{\label{f:appTMG}Bifurcation diagrams for growth on TMG in the presence
of DNA looping: (a)~$\hat{\alpha}_{1}=4$, (b) $\hat{\alpha}_{1}=15$.
The inset in (b) shows a blow-up of the bifurcation curve in the region,
$\alpha_{1}\lesssim3$. }
\end{figure}

To determine the existence of bistability in the presence of DNA looping,
we begin by observing that $\epsilon_{1}(\chi_{1})>0$ whenever $\bar{\delta}_{1}(\chi_{1})>0$.
Hence, it suffices to confine our attention to $\alpha_{1}(\chi_{1})$
and $\delta_{1}(\chi_{1})$. But $\alpha_{1}$ and $\delta_{1}$ are
positive on the interval $0<\chi_{1}<1$ (resp., $1<\chi_{1}<\infty)$
if and only if $\psi_{1}$ and $\psi_{2}$ are positive (resp., negative).
Thus, we are led to consider the signs of $\psi_{1}$ and $\psi_{2}$
on $\chi_{1}>0$.

It follows from the analysis of $\psi_{1}$ that:

\begin{enumerate}
\item If $0<\hat{\alpha}_{1}<(5/3)^{5}$, then $\psi_{1}<0$ for all $\chi_{1}>0$.
\item If $\hat{\alpha}_{1}>(5/3)^{5}$, then $\psi_{1}$ has two positive
roots, $0<\underline{\chi}_{1}<\overline{\chi}_{1}$, and $\psi_{1}>0$
if and only if $\underline{\chi}_{1}<\chi_{1}<\overline{\chi}_{1}$.
Furthermore, $\underline{\chi}_{1},\overline{\chi}_{1}<1$ when $(5/3)^{5}<\hat{\alpha}_{1}<16$,
and $\underline{\chi}_{1}<1<\overline{\chi}_{1}$ when $\hat{\alpha}_{1}>16$.
\end{enumerate}
Similarly, the analysis of $\psi_{2}$ shows that:

\begin{enumerate}
\item If $0<\hat{\alpha}_{1}<1$, then $\psi_{2}<0$ for all $\chi_{1}>0$.
\item If $\hat{\alpha}_{1}>1$, then $\psi_{2}>0$ if and only if $0<\chi_{1}<\chi_{1}^{*}\equiv\hat{\alpha}_{1}^{1/4}-1$.
Furthermore, if $\hat{\alpha}_{1}>(5/3)^{5}$, then $\chi_{1}^{*}$
lies between $\overline{\chi}_{1}$ and 1.
\end{enumerate}
Taken together, these results imply that there are three distinct
types of the bifurcation diagrams:

\begin{enumerate}
\item If $0\le\hat{\alpha}_{1}<(5/3)^{5}$, then $\alpha_{1}(\chi_{1}),\bar{\delta}_{1}(\chi_{1})$
are positive if and only if $\chi_{1}>1$ (Fig.~\ref{f:appTMG}a).
\item If $(5/3)^{5}\le\hat{\alpha}_{1}<16$, then $\bar{\delta}_{1}(\chi_{1}),\alpha_{1}(\chi_{1})$
are positive if and only if $\underline{\chi}_{1}<\chi_{1}<\overline{\chi}_{1}$
or $\chi>1$, where $0<\underline{\chi}_{1}<\overline{\chi}_{1}<1$
are the positive roots of $\alpha_{1}(\chi_{1})$. Each of these two
intervals of existence yields a bistability region (Fig.~\ref{f:appTMG}b).
\item If $\hat{\alpha}_{1}>16$, then $\bar{\delta}_{1}(\chi_{1}),\alpha_{1}(\chi_{1})$
are positive if and only if $\underline{\chi}_{1}<\chi<1$ or $\chi>\overline{\chi}_{1}$,
where $\underline{\chi}_{1}$ and $\overline{\chi}_{1}$ are the positive
roots of $\alpha_{1}(\chi_{1})$ (Fig.~\ref{fig:TMGbdWithLooping}a,b).
\end{enumerate}
The geometry of the cusps in the foregoing figures follow from the
relations\begin{align*}
\frac{d\alpha_{1}}{d\chi_{1}} & =2\frac{\hat{\alpha}_{1}\left(3\chi_{1}^{2}-3\chi_{1}+2\right)+\left(1+\chi_{1}\right)^{5}\left(\chi_{1}-2\right)}{\left(1+\chi_{1}\right)^{3}\left(1-\chi_{1}\right)^{2}},\\
\frac{d\delta_{1}}{d\chi_{1}} & =\frac{\chi_{1}}{\left(1+\chi_{1}\right)^{2}}\frac{d\alpha_{1}}{d\chi_{1}},\end{align*}
which imply that there is a cusp whenever $d\alpha_{1}/d\chi_{1}=0$.
It follows from the properties of $\psi_{1}$ that there are two cusps
if and only if $\left(5/3\right)^{5}<\hat{\alpha}_{1}<16$. If $\hat{\alpha}_{1}>16$,
one of these cusps disappears because its $\alpha_{1}$-coordinate
is negative.

\section{\label{a:LactoseBD}Bifurcation diagram for growth on lactose}

In this case, the bifurcation points satisfy the equations\begin{align*}
g(\epsilon_{1}) & \equiv\frac{1}{1+\alpha_{1}/\left(1+\delta_{1}\epsilon_{1}\right)^{2}+\hat{\alpha}_{1}/\left(1+\delta_{1}\epsilon_{1}\right)^{4}}-\epsilon_{1}^{2}=0,\\
g_{\epsilon_{1}}(\epsilon_{1}) & =\frac{2\alpha_{1}\delta_{1}/\left(1+\delta_{1}\epsilon_{1}\right)^{3}+4\hat{\alpha}_{1}\delta_{1}/\left(1+\delta_{1}\epsilon_{1}\right)^{5}}{\left[1+\alpha_{1}/\left(1+\delta_{1}\epsilon_{1}\right)^{2}+\hat{\alpha}_{1}/\left(1+\delta_{1}\epsilon_{1}\right)^{4}\right]^{2}}-2\epsilon_{1}=0.\end{align*}
which imply that\begin{align*}
\frac{\alpha_{1}}{\left(1+\delta_{1}\epsilon_{1}\right)^{2}}+\frac{\hat{\alpha}_{1}}{\left(1+\delta_{1}\epsilon_{1}\right)^{4}} & =\frac{1}{\epsilon_{1}^{2}}-1,\\
\frac{\alpha_{1}}{\left(1+\delta_{1}\epsilon_{1}\right)^{2}}+2\frac{\hat{\alpha}_{1}}{\left(1+\delta_{1}\epsilon_{1}\right)^{4}} & =\left(1+\frac{1}{\delta_{1}\epsilon_{1}}\right)\frac{1}{\epsilon_{1}^{2}}.\end{align*}
Since $\chi_{1}=\delta_{1}\epsilon_{1}$ , we have two equations that
are linear with respect to $\alpha_{1}$, $\hat{\alpha}_{1}$, and
$1/\epsilon_{1}^{2}$. Solving for $\alpha_{1}$ and $1/\epsilon_{1}$
in terms of $\hat{\alpha}_{1}$ and $\chi_{1}$ yields \begin{align*}
\epsilon_{1}(\chi_{1}) & =\sqrt{\frac{1}{\chi_{1}\left\{ \hat{\alpha}_{1}/\left(1+\chi_{1}\right)^{4}-1\right\} }},\\
\alpha_{1}(\chi_{1}) & =\frac{\psi_{3}(\chi_{1})}{\left(1+\chi_{1}\right)^{2}},\;\psi_{3}(\chi_{1})\equiv\hat{\alpha}_{1}\left(\chi_{1}-1\right)-\left(1+\chi_{1}\right)^{5},\\
\delta_{1}(\chi_{1}) & =\frac{\chi_{1}}{\epsilon(\chi_{1})}=\chi_{1}^{3/2}\sqrt{\frac{\hat{\alpha}_{1}}{\left(1+\chi_{1}\right)^{4}}-1},\end{align*}
For each fixed $\hat{\alpha}_{1}\ge0$, these relations provide a
parametric representation of the bifurcation curve.

In the absence of DNA looping, the bifurcation curve does not exist
because the $\epsilon_{1}$- and $\delta_{1}$-coordinates of the
curve are imaginary (and the $\alpha_{1}$-coordinate is negative)
for all $\chi_{1}$.

In the presence of DNA looping, the bifurcation curve exists for all
$\hat{\alpha}_{1}>5^{5}/2^{4}\approx195$. To see this, observe that
since \begin{equation}
\alpha_{1}(\chi_{1})>0\Rightarrow\hat{\alpha}_{1}\left(\chi_{1}+1\right)>\hat{\alpha}_{1}\left(\chi_{1}-1\right)>\left(1+\chi_{1}\right)^{5},\label{eq:Ineq}\end{equation}
$\epsilon_{1}$ and $\delta_{1}$ are positive whenever $\alpha_{1}$
is positive. Hence, it suffices to confine our attention to $\alpha_{1}$.
Now, $\alpha_{1}>0$ if and only $\psi_{3}>0$. One can solve the
equations $\psi_{3}=d\psi_{3}/d\chi_{1}=0$ to conclude that $\psi_{3}>0$
for some $\chi_{1}>0$ if and only if $\hat{\alpha}_{1}>5^{5}/2^{4}$.
In this case, $\psi_{3}$ and $\alpha_{1}$ have two roots, $0<\underline{\chi}_{1}<\overline{\chi}_{1}$,
and are positive if and only if $\chi_{1}$ lies between these roots.

The bifurcation diagram is qualitatively similar to Fig.~\ref{fig:bdLactose}a
for all $\hat{\alpha}_{1}>5^{5}/2^{4}$ because under this condition,
$\alpha_{1}$ and $\delta_{1}$ are positive on $(\underline{\chi}_{1},\overline{\chi}_{1})$,
and achieve a unique maximum on $(\underline{\chi}_{1},\overline{\chi}_{1})$
at the very same value of $\chi_{1}$. To see this, observe that $\alpha_{1}$
has at least one maximum in $(\underline{\chi}_{1},\overline{\chi}_{1})$.
In fact, it has exactly one maximum because\[
\frac{d\alpha_{1}}{d\chi_{1}}=\frac{\hat{\alpha}_{1}\left(3-\chi_{1}\right)-3\left(1+\chi_{1}\right)^{5}}{\left(1+\chi_{1}\right)^{3}}\]
cannot have more than one zero on this interval. On the other hand,
$\delta_{1}>0$ on $[\underline{\chi}_{1},\overline{\chi}_{1}]$ because
$\delta_{1}>\alpha_{1}$. Furthermore, $\delta_{1}$ and $\alpha_{1}$
attain a maximum at the very same $\chi_{1}$ because\begin{align*}
\frac{d\delta_{1}}{d\chi_{1}} & =\frac{1}{2}\left(\frac{\chi_{1}}{1+\chi_{1}}\right)^{2}\frac{1}{\delta_{1}}\frac{d\alpha_{1}}{d\chi_{1}},\\
\frac{d^{2}\delta_{1}}{d\chi_{1}^{2}} & =\frac{1}{2\delta_{1}}\frac{d^{2}\alpha_{1}}{d\chi_{1}^{2}},\end{align*}
The first relation implies that $d\delta_{1}/d\chi_{1}=0$ precisely
when $d\alpha_{1}/d\chi_{1}=0$. The second relation implies that
when $d\delta_{1}/d\chi_{1}$ (and hence, $d\alpha_{1}/d\chi_{1}$)
is zero, $d^{2}\delta_{1}/d\chi_{1}^{2}$ and $d^{2}\alpha_{1}/d\chi_{1}^{2}<0$
have the same sign.

The width of the cusp-shaped region always increases as one moves
away from the cusp point (Fig.~\ref{fig:bdLactose}a). To see this,
observe that the above equations imply that \[
\frac{d\delta_{1}^{2}}{d\alpha_{1}}=\frac{\chi_{1}}{\left(1+\chi_{1}\right)^{2}}\]
is an increasing function of $\chi_{1}$. Hence, the slope at any
point on the upper branch of the bifurcation curve is always greater
than the slope of any point on the lower branch.

\section{\label{a:GLbd}Bifurcation diagram for growth on lactose + glucose}

In this case, the steady states satisfy the equations\begin{align}
\rho_{1} & =\left(\epsilon_{1}+\alpha\epsilon_{2}\right)\epsilon_{1},\;\rho_{1}\equiv\frac{1}{1+\alpha_{1}/\left(1+\delta_{1}\epsilon_{1}\right)^{2}+\hat{\alpha}_{1}/\left(1+\delta_{1}\epsilon_{1}\right)^{4}},\label{eq:AppGLss1}\\
\rho_{2} & =\left(\epsilon_{1}+\alpha\epsilon_{2}\right)\epsilon_{2},\rho_{2}\equiv\frac{\alpha}{1+\alpha_{2}/\left(1+\delta_{2}\epsilon_{2}\right)},\label{eq:AppGLss2}\end{align}
where $\alpha=\beta\delta_{1}/\delta_{2}$. If a steady state corresponds
to a fold bifurcation point, the determinant of the Jacobian, \[
J=\left[\begin{array}{cc}
\frac{2\alpha_{1}\delta_{1}/\left(1+\delta_{1}\epsilon_{1}\right)^{3}+4\hat{\alpha}_{1}\delta_{1}/\left(1+\delta_{1}\epsilon_{1}\right)^{5}}{\left\{ 1+\alpha_{1}/\left(1+\delta_{1}\epsilon_{1}\right)^{2}+\hat{\alpha}_{1}/\left(1+\delta_{1}\epsilon_{1}\right)^{4}\right\} ^{2}}-2\epsilon_{1}-\alpha\epsilon_{2} & -\alpha\epsilon_{1}\\
-\epsilon_{2} & \frac{\alpha}{\left\{ 1+\alpha_{2}/\left(1+\delta_{2}\epsilon_{2}\right)\right\} ^{2}}\frac{\alpha_{2}\delta_{2}}{\left(1+\delta_{2}\epsilon_{2}\right)^{2}}-\epsilon_{1}-2\alpha\epsilon_{2}\end{array}\right],\]
 at that steady state must also be zero. Now, it follows from (\ref{eq:AppGLss1})
that at a steady state \begin{align*}
J_{11} & =\left(\epsilon_{1}+\alpha\epsilon_{2}\right)\epsilon_{1}\frac{2\alpha_{1}\delta_{1}/\left(1+\delta_{1}\epsilon_{1}\right)^{3}+4\hat{\alpha}_{1}\delta_{1}/\left(1+\delta_{1}\epsilon_{1}\right)^{5}}{1+\alpha_{1}/\left(1+\delta_{1}\epsilon_{1}\right)^{2}+\hat{\alpha}_{1}/\left(1+\delta_{1}\epsilon_{1}\right)^{4}}-2\epsilon_{1}-\alpha\epsilon_{2}\\
 & =\left(\epsilon_{1}+\alpha\epsilon_{2}\right)p-\epsilon_{1}\end{align*}
where\[
p\equiv2\frac{\delta_{1}\epsilon_{1}}{1+\delta_{1}\epsilon_{1}}\frac{\alpha_{1}/\left(1+\delta_{1}\epsilon_{1}\right)^{2}+2\hat{\alpha}_{1}/\left(1+\delta_{1}\epsilon_{1}\right)^{4}}{1+\alpha_{1}/\left(1+\delta_{1}\epsilon_{1}\right)^{2}+\hat{\alpha}_{1}/\left(1+\delta_{1}\epsilon_{1}\right)^{4}}-1.\]
Similarly, (\ref{eq:AppGLss2}) implies that at a steady state \begin{align*}
J_{22} & =\left(\epsilon_{1}+\alpha\epsilon_{2}\right)\epsilon_{2}\frac{1}{1+\alpha_{2}/\left(1+\delta_{2}\epsilon_{2}\right)}\frac{_{2}}{\left(1+\delta_{2}\epsilon_{2}\right)^{2}}-\epsilon_{1}-2\alpha\epsilon_{2}\\
 & =\left(\epsilon_{1}+\alpha\epsilon_{2}\right)q-\alpha\epsilon_{2},\end{align*}
where\[
q\equiv\frac{\alpha_{2}}{\alpha_{2}+\left(1+\delta_{2}\epsilon_{2}\right)}\frac{\delta_{2}\epsilon_{2}}{1+\delta_{2}\epsilon_{2}}-1.\]
It follows that $\det J$ is zero at a steady state if and only if\[
\left(\epsilon_{1}+\alpha\epsilon_{2}\right)\left[\left(\epsilon_{1}+\alpha\epsilon_{2}\right)pq-\alpha p\epsilon_{2}-q\epsilon_{1}\right]=0,\]
i.e.,\begin{equation}
\alpha=\frac{\epsilon_{1}}{\epsilon_{2}}h,\; h\equiv\frac{1/p-1}{1-1/q}.\label{eq:AppGLdet}\end{equation}
The bifurcation points satisfy eqs.~(\ref{eq:AppGLss1})--(\ref{eq:AppGLdet}).

To determine the parametric representation of the bifurcation points,
observe that (\ref{eq:AppGLss1})--(\ref{eq:AppGLss2}) yield $\epsilon_{1}/\epsilon_{2}=\rho_{1}/\rho_{2},$
which can be substituted in (\ref{eq:AppGLdet}) to obtain $\alpha=\left(\rho_{1}/\rho_{2}\right)h$,
i.e., \begin{equation}
\alpha=\frac{\beta\delta_{1}}{\delta_{2}}=\sqrt{\frac{1+\alpha_{2}/\left(1+\chi_{2}\right)}{1+\alpha_{1}/\left(1+\chi_{1}\right)^{2}+\hat{\alpha}_{1}/\left(1+\chi_{1}\right)^{4}}}\sqrt{h},\label{eq:AppalphaParam}\end{equation}
where $h$ is now a function of $\chi_{1}$ and $\chi_{2}$. Eqs.~(\ref{eq:AppGLss1})--(\ref{eq:AppGLss2})
also imply that\[
\rho_{1}-\alpha\rho_{2}=\epsilon_{1}^{2}\left[1-\alpha^{2}\left(\frac{\epsilon_{2}}{\epsilon_{1}}\right)^{2}\right]=\epsilon_{1}^{2}\left[1-\alpha^{2}\left(\frac{\rho_{2}}{\rho_{1}}\right)^{2}\right],\]
whence\[
\epsilon_{1}^{2}=\rho_{1}\frac{1}{1+\alpha\left(\rho_{2}/\rho_{1}\right)}=\rho_{1}\frac{1}{1+h}.\]
Hence \[
\epsilon_{1}=\sqrt{\frac{1}{1+\alpha_{1}/\left(1+\chi_{1}\right)^{2}+\hat{\alpha}_{1}/\left(1+\chi_{1}\right)^{4}}}\sqrt{\frac{1}{1+h}},\]
and \begin{equation}
\delta_{1}(\chi_{1},\chi_{2})=\frac{\chi_{1}}{\epsilon_{1}}=\chi_{1}\sqrt{1+\alpha_{1}/\left(1+\chi_{1}\right)^{2}+\hat{\alpha}_{1}/\left(1+\chi_{1}\right)^{4}}\sqrt{1+h}.\label{eq:Appdelta1Param}\end{equation}
Finally,\[
\epsilon_{2}=\frac{\rho_{2}}{\rho_{1}}\epsilon_{1}=\frac{\alpha\sqrt{1+\alpha_{1}/\left(1+\chi_{1}\right)^{2}+\hat{\alpha}_{1}/\left(1+\chi_{1}\right)^{4}}}{1+\alpha_{2}/\left(1+\chi_{2}\right)}\sqrt{\frac{1}{1+h}}.\]
It follows from (\ref{eq:AppalphaParam}) that \begin{align*}
\epsilon_{2} & =\frac{1}{\sqrt{1+\alpha_{2}/\left(1+\chi_{2}\right)}}\sqrt{\frac{h}{1+h}},\end{align*}
and\begin{equation}
\delta_{2}(\chi_{1},\chi_{2})=\frac{\chi_{2}}{\epsilon_{2}}=\chi_{2}\sqrt{1+\alpha_{2}/\left(1+\chi_{2}\right)}\sqrt{\frac{1+h}{h}}.\label{eq:Appdelta2Param}\end{equation}
Substituing (\ref{eq:Appdelta1Param})--(\ref{eq:Appdelta2Param})
in (\ref{eq:AppalphaParam}) yields\begin{equation}
\beta=\frac{\chi_{1}}{\chi_{2}}h\Leftrightarrow\beta\chi_{2}\frac{\alpha_{2}\left(\chi_{2}+2\right)+2\left(1+\chi_{2}\right)^{2}}{\alpha_{2}+\left(1+\chi_{2}\right)^{2}}=\frac{1}{p(\chi_{1})}-1.\label{eq:AppBeta}\end{equation}
Eqs.~(\ref{eq:Appdelta1Param})--(\ref{eq:AppBeta}) provide a parametric
representation of the bifurcation points.

\begin{figure}
\noindent \begin{centering}\subfigure[]{\includegraphics[width=3in,keepaspectratio]{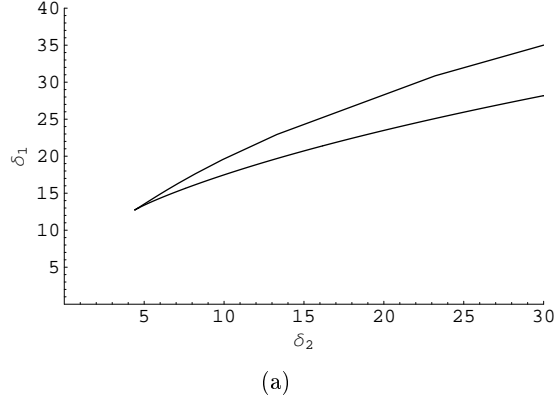}}\par\end{centering}

\caption{\label{f:GLbdPlotNoLooping}The bifurcation diagram for growth on
glucose and lactose at $\alpha_{1}=170$, $\hat{\alpha}_{1}=0$, $\alpha_{2}=4$,
$\beta=1$). The $\delta_{2}$-coordinate of the bifurcation curve
is always positive, i.e., bistability is infeasible at sufficiently
small $\delta_{2}$. This reflects the fact that if the repression
due to DNA looping is small, bistability is impossible during growth
on pure lactose ($\delta_{2}=0$).}
\end{figure}

Bistability is feasible if and only if and only if $0<p<1$ for some
$\chi_{1}>0$. To see this, observe that the LHS of (\ref{eq:AppBeta})
is a monotonically increasing function of $\chi_{2}$ for all $\chi_{2}\ge0$,
and the RHS of (\ref{eq:AppBeta}) is positive if and only if $0<p<1$.
It follows that for every $\chi_{1}>0$ such that $0<p(\chi_{1})<1$,
there is a unique $\chi_{2}>0$ satisfying (\ref{eq:AppBeta}). These
pairs, $\chi_{1},\chi_{2}>0$, define a curve on the $\chi_{1}\chi_{2}$-plane,
and the variation of $\delta_{1}(\chi_{1},\chi_{2})$ and $\delta_{2}(\chi_{1},\chi_{2})$
along this curve delineates the bistable region on the $\delta_{1}\delta_{2}$-plane.
In contrast, for every $\chi_{1}>0$ such that $p(\chi_{1})<0$ or
$p(\chi_{1})>1$, the RHS of (\ref{eq:AppBeta}) is negative. Hence,
there is no $\chi_{2}>0$ satisfying (\ref{eq:AppBeta}), and no bistability.

The condition, $p>0$ for some $\chi_{1}>0$, is satisfied if and
only if $\alpha_{1}$ and $\hat{\alpha}_{1}$ lie above the full curve
shown in Fig.~\ref{f:bdA1A2H}. To see this, observe that $p$ has
a maximum on $\chi_{1}>0$ since $p(0)=-1$, $p(\infty)=-1$, and
$p(\chi_{1})>-1$ for $0<\chi_{1}<\infty$. The value of $\alpha_{1}$
and $\hat{\alpha}_{1}$ at which a maximum of $p$ touches the $\chi_{1}$-axis
satisfies the equations\begin{align*}
p(\chi_{1})=0 & \Leftrightarrow\left(\chi_{1}-1\right)\frac{\alpha_{1}}{\left(1+\chi_{1}\right)^{2}}+\left(3\chi_{1}-1\right)\frac{\hat{\alpha}_{1}}{\left(1+\chi_{1}\right)^{2}}=1+\chi_{1},\\
p_{\chi_{1}}(\chi_{1})=0 & \Leftrightarrow\left(3-\chi_{1}\right)\frac{\alpha_{1}}{\left(1+\chi_{1}\right)^{2}}+\left(7-9\chi_{1}\right)\frac{\hat{\alpha}_{1}}{\left(1+\chi_{1}\right)^{2}}=1+\chi_{1}\end{align*}
which can be solved to obtain\[
\alpha_{1}(\chi_{1})=2\frac{\left(1+\chi_{1}\right)^{3}\left(3\chi_{1}-2\right)}{3\chi_{1}^{2}-3\chi_{1}+2},\;\hat{\alpha}_{1}(\chi_{1})=\frac{\left(1+\chi_{1}\right)^{5}\left(2-\chi_{1}\right)}{3\chi_{1}^{2}-3\chi_{1}+2}.\]
The above relations define the full curve shown in Fig.~\ref{f:bdA1A2H}.
Evidently, the $\hat{\alpha}_{1}$- and $\alpha_{1}$-intercepts of
the curve are 27 and $(5/3)^{5}$, respectively. Moreover, since $\partial p/\partial\hat{\alpha}_{1}>0$
for all $\chi_{1}>0$, the condition, $0<p$ for some $\chi_{1}>0$,
is satisfied precisely when $\alpha_{1}$ and $\hat{\alpha}_{1}$
lie above this curve.

If $\alpha_{1}$ and $\hat{\alpha}_{1}$ lie above the full curve
in Fig.~\ref{f:bdA1A2H}, there is an interval, say, $(\underline{\chi}_{1},\overline{\chi}_{1})$,
on which $p>0$. But there are two possibilities. Either $p$ never
exceeds 1 on this interval, or it exceeds 1 on some subinterval of
$(\underline{\chi}_{1},\overline{\chi}_{1})$, say, $(\chi_{1,l},\chi_{1,u})$.
In the first case, the $\chi_{2}$-coordinate of the curve defined
by (\ref{eq:AppBeta}), and hence, the $\delta_{2}$-coordinate of
the bifurcation curve is always positive, i.e., the bifurcation diagram
has the form shown in Fig.~\ref{f:GLbdPlotLooping}. In the second
case, the $\chi_{2}$-coordinate of the curve defined by (\ref{eq:AppBeta})
is zero at $\chi_{1}=\chi_{1,l},\chi_{1,u}$, and the $\delta_{2}$-coordinate
of the bifurcation curve is zero at the corresponding point. The bifurcation
diagram therefore has the form shown in Fig.~\ref{f:GLbdPlotLooping}a.

One can check (by the method similar to the one shown above) that
$p$ exceeds 1 for some $\chi_{1}>0$ if and only $\alpha_{1}$ and
$\hat{\alpha}_{1}$ lie above the curve defined by the relations\[
\alpha_{1}(\chi_{1})=2\frac{\left(1+\chi_{1}\right)^{3}\left(2\chi_{1}-3\right)}{3-\chi_{1}},\;\hat{\alpha}_{1}(\chi_{1})=3\frac{\left(1+\chi_{1}\right)^{5}}{3-\chi_{1}},\]
which is shown as the dashed curve in Fig.~\ref{f:bdA1A2H}. Evidently,
the $\hat{\alpha}_{1}$- intercept of the curve is $5^{5}/2^{4}$.

\end{document}